\documentclass[aps,prd,preprint,tightenlines,groupedaddress,showpacs]{revtex4}
\usepackage{epsf,epsfig,graphics,graphicx}
\begin{document}
\title{ Nonequilibrium dynamics of moving mirrors in quantum  fields: Influence
functional and Langevin equation}
\author{Chun-Hsien Wu}
\email{chunwu@phys.sinica.edu.tw}
\author{Da-Shin Lee}
\email{dslee@mail.ndhu.edu.tw}
\affiliation{Department of Physics,
National Dong-Hwa University, Hualien, Taiwan, R.O.C.}

\date{\today}

\begin{abstract}
We employ the Schwinger-Keldysh formalism to study the
nonequilibrium dynamics of the mirror with perfect reflection
moving in a quantum field.  In the case where the mirror undergoes the small
displacement, the coarse-grained effective action is obtained
 by integrating out the quantum field with the method of influence functional.
 The semiclassical Langevin equation is derived,  and  is found to
 involve two levels of backreaction effects  on the dynamics of
mirrors:  radiation reaction induced by the motion of the mirror
and backreaction dissipation  arising from fluctuations in quantum
field via a fluctuation-dissipation relation.  Although the corresponding
theorem of fluctuation and dissipation for the case with the small mirror's displacement
is of model independence, the study from the first principles
derivation shows that the  theorem is also {\it
independent} of the regulators introduced to deal with
short-distance divergences from the quantum field.
 Thus, when the method of  regularization is
introduced to compute the dissipation and fluctuation effects,
this theorem must be fulfilled as the results are obtained by
taking the short-distance limit in the end of calculations. The
backreaction effects from vacuum fluctuations on moving mirrors
are found to be hardly detected while those effects from thermal
fluctuations may be detectable.

\end{abstract}

 \maketitle

\section{Introduction}

Zero-point  fluctuations   due to the imposition of the boundary
conditions can lead to an impact on macroscopic physics. One of
the most celebrated examples is the attractive Casimir force
between two parallel conducting plates~\cite{CA}.  However, the
dynamics of  fluctuations subject to the moving boundary
  may also be detectable,
sometimes  referred to as the dynamical Casimir
effects~\cite{HU1,HU2,HU3,HU4,JR1,FD,FV,JR2}. Consider a perfectly
reflecting mirror moving in quantum fields. The boundary
conditions on quantum fields corresponding to perfect reflection
result in the interaction of the mirror with the fields. The
motion of the mirror, which leads to the moving boundary, can
create quantum radiation that in turn damps out the motion of the
mirror as a result of the motion-induced radiation reaction force.
In fact, as required by Lorentz invariance of quantum fields, this
radiation reaction force vanishes for a motion with uniform
velocity. In a motion of uniform acceleration, the mirror suffers
from the same fluctuations as if it was at rest in a thermal bath
due to the Unruh effects~\cite{UN}, also leading to the zero
dissipative radiation reaction  force. Fulling and Davies have
computed this force for a moving mirror in a massless scalar field
in the 1+1 dimensional spacetime. It turns out that the induced
dissipative force is proportional to the third time derivative of
the mirror's position~\cite{FD}. In 3+1 dimensional spacetime, the problem
has been studied by Ford and Vilenkin in terms of a first order approximation of the mirror's
displacement. The corresponding dissipative force then is given by
the fifth time derivative of the position in the non-relativistic
limit~\cite{FV}. However, as we know,  all quantum fields exhibit
fluctuations that manifest themselves through the fluctuating
forces on the mirror such as fluctuations of Casimir
forces~\cite{BA,WF,FS,MMF}. Thus, through a fluctuation and
dissipation relation as in the case of Brownian motion, in
addition to  motion-induced radiation reaction, the mirror must
experience the backreaction dissipation effect arising from the
force fluctuations~\cite{HU1,HU2,HU3,HU4}. In this paper, a first
principles derivation is provided to study the dynamics of the
moving mirror by taking account of the backreaction effects from
quantum fields consistently within the context of the
Schwinger-Keldysh formalism. Coarse-graining the degrees of
freedom of quantum fields results in  the coarse-grained effective
action with the method of influence functional. This approach can
naturally lead to the Langevin equation in the semiclassical
approximation, and allows us to obtain the corresponding
fluctuation and dissipation theorem from a microscopic point of
view.

The problem addressed in this paper can be viewed as a special
case of the larger problem of radiation reaction arising from
vacuum or/and thermal fluctuations~\cite{MS,CT,FLO}. Especially,
in vacuum, this problem can probe the nature of vacuum
fluctuations and viscosity in relation with the backreaction of
cosmological particle creation.

 This paper is organized
as follows: The theory to describe the
interaction between the mirror and quantum fields is discussed in
Section II. In Section III  the Langevin equation in the
semiclassical approximation is obtained by integrating out quantum
fields. The backreaction forces  are computed in Section VI. In
Section V we derive the corresponding fluctuation and dissipation
theorem, and discuss its applications. The calculations to obtain
the dynamics of the moving mirror involving backreaction effects
from quantum fields in vacuum and at finite temperature
respectively are presented in Section VI. We then draw the
conclusions in Section VII.

\section{Field perturbation driven by small mirror's displacement}

We consider a mirror with perfect reflection moving in a quantum
field given by a massless, minimally coupled  scalar field. As a
result, the corresponding boundary condition on the scalar field
is as follows:
\begin{equation}
\phi \mid_{\rm S} =0 \, . \label{boundcond-general}
\end{equation}
 The mirror of mass $m$ and area $A$ is oriented parallel to the $z=0$ plane.
We assume that the mirror has a small displacement $\delta q(t)$
along the $z$-direction from the origin  which can be obtained,
for example  by applying the classical external force. Then,  the
boundary condition above can be expressed in the specific form:
\begin{equation}
\phi (x,y,\delta q (t),t)=0 \, . \label{boundcond}
\end{equation}
 To first order in $\delta q(t)$, we obtain
\begin{equation}
[\phi (x,y,0,t)+\delta q(t)\,  \partial_z \phi(x,y,0,t)+ \cdot
\cdot \cdot ]=0 \, . \label{boundcond-perturb}
\end{equation}
We then further assume that  the mirror's surface  $S$ has small perturbations
induced from the motion of the mirror. This means that the quantum
field $\phi$ can be written as:
\begin{equation}
\phi=\phi_0 + \delta \phi \, , \label{fieldperturb}
\end{equation}
where the field $\phi_0$ corresponds to the field fluctuations
with respect to  the unperturbed  surface $S_0$ at the $ z=0$
plane, while the field  $\delta \phi$ is   the induced
fluctuations on the surface $S$ driven by the motion of the
mirror, and is of order $\delta q(t)$. Thus, together with Eqs.(\ref{boundcond-perturb}),
(\ref{fieldperturb}), and   the vanishing boundary condition  of
the field $\phi_0$ on $S_0$,
\begin{equation}
\phi_0 (x,y,0) =0 \, , \label{fieldbound}
\end{equation}
the perturbed  field  $\delta \phi$, to first order in $\delta
q(t)$, is  given by:
\begin{equation}
\delta \phi (x,y,0,t)=- \delta q(t) \, \partial_z \phi_0 (x,y,0,t)
\,. \label{perturbedfield}
\end{equation}
The  force acting on both sides of the mirror is given by the area
integral  of the $z-z$ component of the stress tensor in terms of
field operators:
\begin{equation}
F(t)= F(0^-,t)-F(0^+,t) = \int_{A} dx\, dy\, \left[ T_{zz} ( x,y,
0^{-},t)-T_{zz} ( x,y, 0^{+},t) \right] \, , \label{force1}
\end{equation}
where
\begin{equation}
T_{zz}= \frac{1}{2} \left[ \left(\partial_t \phi \right)^2
+\left(\partial_z \phi \right)^2-\left(\partial_x \phi \right)^2-
\left(\partial_y \phi \right)^2 \right] \, .
\end{equation}
To first order in $\delta q(t)$, we can write
\begin{equation}
T_{zz}= T_{0, zz} +\delta T_{zz} \, ,
\end{equation}
where
\begin{eqnarray}
T_{0,zz}&=& \frac{1}{2} \left[ \left(\partial_t \phi_0 \right)^2
+\left(\partial_z \phi_0 \right)^2-\left(\partial_x \phi_0
\right)^2- \left(\partial_y \phi_0 \right)^2 \right] \, ,
\label{stress0}
\\
\delta T_{zz} &=&\ \frac{1}{2} \,  \left[
\partial_t \phi_0 \,
\partial_t \delta \phi +\partial_t  \delta \phi \, \partial_t  \phi_0 +
\partial_z \phi_0 \,
\partial_z \delta \phi +   \partial_z \delta \phi \, \partial_z  \phi_0 \right. \nonumber \\
&& \left. -
\partial_x \phi_0 \,
\partial_x \delta \phi - \partial_x \delta \phi \, \partial_x
\phi_0 -  \partial_y  \phi_0 \, \partial_y \delta \phi  -
\partial_y \delta \phi \,
\partial_y  \phi_0 \right] \, . \label{stress}
\end{eqnarray}
It leads to the following effective force term:
\begin{equation}
F(t)= F_0 (t) +\delta q(t) \, \frac{\delta F}{\delta q}(t) \, ,
\label{force}
\end{equation}
where
\begin{eqnarray}
F_0 (t)&=& F_0 (0^-,t)-F_0 (0^+,t)  = \int_A dx \, dy \, \left[
T_{0,zz}
(x,y,0^-,t) - T_{0, zz} (x,y,0^+,t) \right] \, , \label{forcestress0} \\
\delta q (t) \frac{ \delta F}{\delta q} (t)&=&  \delta q (t)
\left[ \frac{ \delta F}{\delta q} (0^-, t)+\frac{ \delta F}{\delta
q} (0^+,t) \right] =\delta q (t) \int_A dx \, dy \,
\left[\frac{\delta T_{zz}}{\delta q} (x,y,0^-,t)+ \frac{\delta
T_{zz}}{\delta q}( x,y,0^+,t) \right] \, . \label{forcestress} \nonumber \\
\end{eqnarray}
Notice that the $+$ sign for the perturbed force in Eq.(\ref{forcestress}) is due to the fact that there is a sign difference for the
mirror's displacement seen from the forces in the opposite sides
of the mirror. The $ \delta q $ in Eq.(\ref{forcestress}) is
defined to be the mirror's displacement with respect to the force
from $ z=0^-$.

Thus,
the  Lagrangian can be expressed as:
\begin{equation}
L[\delta q, \phi_0 ]= \frac{1}{2} m (\delta {\dot q} )^2- V
(\delta q ) + \, \delta q(t) F_0 (t) + \frac{1}{2}  \,\delta q^2
(t) \frac{\delta F}{\delta q}(t) + \int d^3 {\bf x} \, \left[
\frac{1}{2} (\partial_{t} \phi_0 )^2- \frac{1}{2} ( {\vec
\bigtriangledown} \phi_0 )^2 \right] \, , \label{lagrangian}
\end{equation}
which is subject to the boundary condition on the field $\phi_0$
given by:
\begin{equation}
\phi_0 ( x,y,0,t)=0 \,. \label{bondcon_0}
\end{equation}
The classical external force is also considered to apply to
the mirror with potential energy $ V(\delta q)$. Units with $
\hbar=c=1$ are used, and factors $ \hbar $ and $c$ will be
restored in our main results.

Notice that the first term in the right hand side of
Eq.(\ref{force}) given by the homogeneous background scalar field
is evaluated at the unperturbed surface $S_0$, where by symmetry,
the mean pressure force vanishes  as the forces are cancelled from
both sides of the mirror, i.e.,
\begin{equation}
\langle F_0 (t) \rangle =\langle  F_0 (0^-,t \rangle )- \langle
F_0 (0^+,t) \rangle =0 \, . \label{force0}
\end{equation}
However, this force  undergoes fluctuations about its  mean value
due to quantum and/or thermal effects, and will influence the
dynamics of the mirror. In addition,  $ \langle \delta F/\delta q
\rangle \delta q(t)$ is the force arising from the motion of the
mirror.  This motion-induced radiation reaction force has been
extensively studied in the case of  the background scalar field in
vacuum as well as in thermal equilibrium respectively
\cite{FD,FV,MMF}.

Here we employ the Schwinger and Keldysh formalism to obtain the
influence functional on the moving mirror by integrating out the
scalar field with the Lagrangian given by Eq.(\ref{lagrangian}).
Recall that Eq.(\ref{perturbedfield}) is based upon the fact that
the mirror undergoes the small displacement  where the Lagrangian
in Eq.(\ref{lagrangian}) is correct up to order $ {\cal{O}} (
\delta q^2 ) $.
We then implement the semiclassical approximation
 by assuming that the quantum fluctuations coming from the mirror itself can be ignored to obtain its semiclassical Langevin equation.
 It
can be justified by the fact that the typical size of the mirror
is much larger than its compton wavelength. Under this
semiclassical approximation, the dynamics of the mirror is
governed by the coarse-grained effective action involving the
influence functional. However, for  a general interacting field
theory, one cannot obtain the influence functional that includes
all of the quantum loop effects by integrating out the scalar
field. However, here we obtain the influence functional including
all quantum effects up to order  $ {\cal{O}} ( \delta q^2 ) $
consistent with the approximation in the Lagrangian (
Eq.(\ref{lagrangian})) we mention above. It is then expected that
in addition to the classical dynamical equation  given by the
external potential $ V(\delta q)$, the obtained semiclassical
Langevin equation will involve the backreaction force terms
arising from the quantum effects of the scalar field where the
terms are kept up to $ \delta q $. The terms we ignore are say, $
\delta \dot q^2, \delta q \delta \ddot q $, so on and so forth. On
top of that, the noise force with the Gaussian correlation
function  will be introduced to mimic the stochastic dynamics from
the scalar field fluctuations.

\section{ Influence functional and   Langevin equation}

We now consider the case where an initial density  matrix for the
mirror plus the scalar field  at $t=t_i$ is factorized as:
\begin{equation}
{\hat \rho} (t_i) = {\hat \rho}_{\rm mirror} (t_i) \otimes {\hat
\rho}_{\phi_0} (t_i) \, , \label{initialcond}
\end{equation}
where we have  assumed that the mirror and the scalar field are
initially uncoupled. The mirror initially is  assumed to be in its
position eigenstate with the eigenvalue $ \delta q_i$ given by:
\begin{equation}
{\hat \rho}_{\rm mirror} (t_i) = \mid \delta q_i,t_i \rangle \, \langle
\delta q_i,
t_i \mid \, . \label{initialcondmirror}
\end{equation}
However, the scalar field is  in thermal equilibrium  at
temperature $ T=1/\beta $ with the density matrix:
\begin{equation}
{\hat \rho}_{\phi_0} (t_i) =e^{-\beta  H_{\phi_0} } \, ,
\label{initialcondphi}
\end{equation}
where $ H_{\phi_0} $ is the Hamiltonian for the free scalar field
given from Eq.(\ref{lagrangian}). The zero-temperature limit
corresponding to the initial vacuum state for the scalar field can
 be studied  by  taking $ T \rightarrow 0 $. The
interaction between the mirror and the scalar field is  considered
to switch on at $t=t_i$. Then, in the Schrodinger picture,
the density matrix evolves in time as:
\begin{equation}
{\hat \rho} (t_f) = U(t_f, t_i) \, {\hat \rho} (t_i) \, U^{-1}
(t_f, t_i )
\end{equation}
with $ U(t_f,t_i) $, the time evolution operator. Thus, the
nonequilibrium partition function can be defined as:
\begin{equation}
{\cal Z} =  {\rm Tr} \left( U(t_f,t_i) \, {\hat \rho} (t_i) \,
U^{-1} (t_f,t_i) \right) \, .
\end{equation}
We then insert an identity in terms of  a complete set of the
mirror plus field eigenstates,
\begin{equation}
\int \, d q \, d  \phi \mid q, \phi \rangle \, \langle q, \phi
\mid \, =1 \, ,
\end{equation}
between all time evolution operators where  the mirror plus field
state denoted as $ \mid q,\phi \rangle$ is given by  the direct
product of the state of the mirror and that of the scalar field,
namely, $ \mid q,\phi \rangle = \mid q \rangle \otimes \mid \phi
\rangle $. Then, the nonequilibrium partition function becomes
\begin{eqnarray}
{\cal Z} &=& \int \, d \delta q_1 \, d \phi_1 \, \int \, d \delta
q_2 \, d \phi_2 \, \int \, d \delta q_3 \, d \phi_3 \, \langle
\delta q_1, \phi_1 \mid U(t_f,t_i) \ \mid \delta q_2, \phi_2
\rangle  \, \langle \delta q_2, \phi_2 \mid {\hat \rho} (t_i) \mid
\delta q_3,
\phi_3 \rangle\,\nonumber \\
&&  \langle \delta q_3, \phi_3 \mid U^{-1} (t_f,t_i) \mid \delta
q_1, \phi_1 \rangle
\nonumber \\
&=& \int \, d \delta q_1 \, \int {\cal D} \delta  q^+ {\cal D} \delta q^-
  \,\int \, d \phi_1  d \phi_2 \,
  d \phi_3 \,
  \int  {\cal D}
\phi^+_0 {\cal D} \phi^-_0  \,\, \exp  \left\{ i \int_{t_i}^{t_f}
dt \left[ L[\delta q^+,\phi^+_0] -L[\delta q^-, \phi^-_0] \right]
\right\}
\nonumber \\
&& \times \langle  \phi_2 \mid {\hat \rho}_{\phi_0} (t_i) \mid
\phi_3 \rangle\ \, ,
\end{eqnarray}
with the boundary conditions: $ \phi_0^+ ( {\bf x}, t_f )=\phi_0^-
({\bf x},t_f) =\phi_1 ({\bf x})$ , \,  $\phi_0^+ ( {\bf x}, t_i
)=\phi_2 ({\bf x})$, \,  $\phi_0^- ( {\bf x}, t_i )=\phi_3 ({\bf
x})$ as well as $ \delta q^+ (t_f)=\delta q^- (t_f) =\delta q_1
$, \, $\delta q^+ (t_i)=\delta q^- (t_i) =\delta q_i $. This
method for studying nonequilibrium phenomena has been developed by
Schwinger and Keldysh~\cite{SK}. In recent years, it has been
applied in particle physics and cosmology by one of
us~\cite{LN1,LN2}.

Then, we can obtain the coarse-grained effective action from the
nonequilibrium partition function
\begin{equation}
{\cal Z} = \int d q_1 \int {\cal D} \delta q^+ {\cal D} \delta
q^- \exp
i S [\delta q^+, \delta q^-]
\end{equation}
that
involves the influence functional ${\cal F} \left[ \delta q^+,
\delta q^- \right] $   by integrating out  the degrees of freedom
of the scalar field given by:
\begin{equation}
S [\delta q^+, \delta q^-]  =  \left\{ \left[ \frac{1}{2}m (\delta
{\dot q}^+ )^2 - V( \delta q^+ ) \right] - \left[ \frac{1}{2}m
(\delta {\dot q}^-)^2- V(\delta q^-)  \right] \right\}  -i
\ln{\cal F} \left[ \delta q^+, \delta q^- \right] \, .
\label{noneqaction}
\end{equation}
In the semiclassical approximation where we  ignore the  quantum
fluctuations from the mirror itself, the dynamics of the mirror is
governed by the above coarse-grained effective action $ S [\delta
q^+, \delta q^-] $.

To obtain the influence functional,  we now construct the
real-time Green's functions for the scalar field $\phi_0$ with the
boundary condition  in Eq. ({\ref{bondcon_0}). The field can be
expanded in terms of the creation and annihilation operators which
obey the commutation relation with the proper choice of the mode
functions:
\begin{eqnarray}
\phi_0 ({\bf x},t) = \int \frac{d k_{\perp}}{(2 \pi)}  \,\int
\frac{d^2 {\bf k}_{\parallel}}{(2 \pi)^2} && \frac{i \sin
(k_{\perp} z)}{\sqrt k}  \left[ \left( a_{\bf k}  e^{-i k t} +
a^{\dagger}_{-{\bf k}}
e^{i k t} \right) \Theta (z) \right. \nonumber \\
&& \left. + \left( b_{\bf k} e^{-i k t} + b^{\dagger}_{-{\bf k} }
e^{i k t} \right) \Theta(-z) \right] \, e^{i {\bf k}_{\parallel}
\cdot {\bf x}_{\parallel}}  \label{fieldexpand}
\end{eqnarray}
with $ {\bf x} =( {\bf x}_{\parallel}, z) $, and $ {\bf k} =( {\bf
k}_{\parallel}, k_{\perp}) $,  $ k= \mid {\bf k} \mid $ for a
massless scalar field. We have assumed that the area of the mirror $A$ is large
as compared with the relevant length scales under consideration so that the scalar field
can be expanded with respect to an infinite area. However, the area $A$ can be obtained
as the quantum effects of the scalar field on the mirror are included from all over the
 mirror's surface. As we will see, the results we define to measure are in general for per
 unit area. The mirror of perfect reflection, which is
thus of impermeability to  the quantum scalar field, means that
the fluctuations from  opposite  sides of the mirror have no
correlation, thus leading to the commutability between $a_{\bf k},
a^{\dagger}_{{\bf k}}$ and $b_{\bf k}, b^{\dagger}_{{\bf k}}$. The
essential ingredients to perturbative calculations are the
following Green's functions where $ {\bf x}, {\bf x'} $ are in the
same side of the mirror:
\begin{eqnarray}
G^{++}_{0} ({\bf x},{\bf x'}; t,t')&=& G^{>}_{0} ({\bf x},{\bf
x'};t,t') \, \Theta
(t-t') + G^{<}_{0} ({\bf x},{\bf x'}; t,t') \, \Theta (t'-t) \, , \nonumber
\\
G^{--}_{0} ({\bf x},{\bf x'};t,t')&=& G^{>}_{0} ({\bf x},{\bf
x'};t,t') \, \Theta
(t'-t) + G^{<}_{0} ({\bf x},{\bf x'};t,t') \, \Theta (t-t') \, , \nonumber
\\
G^{+-}_{0} ({\bf x},{\bf x'};t,t')&=& G^{<}_{0} ({\bf x},{\bf x'};t,t')  \,
, \nonumber \\
G^{-+}_{0} ({\bf x},{\bf x'};t,t')&=& G^{>}_{0} ({\bf x},{\bf x'};t,t')  \,
; \nonumber \\
G^{>}_{0} ({\bf x},{\bf x'};t,t')&=&  \langle \phi_0 ({\bf x},t)
\, \phi_0 ({\bf x'},t') \rangle  = {\rm Tr} \left({\hat
\rho}_{\phi}\, \phi_0 ({\bf
x},t) \, \phi_0 ({\bf x'},t') \right) \, , \nonumber \\
G^{<}_{0} ({\bf x},{\bf x'};t,t')&=&  \langle \phi_0 ({\bf x}',t')
\, \phi_0 ({\bf x},t) \rangle = {\rm Tr} \left(
{\hat\rho}_{\phi}\, \phi_0 ({\bf x'},t') \, \phi_0 ({\bf x},t)
\right) \, .
\end{eqnarray}
Using the field expansion in Eq.(\ref{fieldexpand}), the Green's
functions  can be expressed as:
\begin{eqnarray}
G^{>}_{0} ({\bf x},{\bf x'};t,t') &=&  G^{>} ({\bf x}-{\bf x'};t-t')
- G^{>} ({\bf x}-{\bf {\bar x}'};t-t') \, , \nonumber \\
  G^{<}_{0} ({\bf x}, {\bf x'};t, t') &=&  G^{<} ({\bf x}-{\bf
  x'};t-t')- G^{<} ({\bf x}-{\bf {\bar x}'};t-t') \, ,
  \label{g0g1}
\end{eqnarray}
where the Green's functions in the right hand side of the above
expressions are the corresponding functions in free space
given by
\begin{eqnarray}
&& G^{>} ({\bf x}-{\bf x'};t-t') = \int \frac{ d^3 {\bf k}}{ (
2\pi)^3} \langle \phi_{\bf k} (t) \phi_{-\bf k} (t') \rangle \,
e^{i {\bf
k}\cdot ({\bf x}-{\bf x'})} \,  \nonumber \\
&=& \int \frac{ d^3 {\bf k}}{ ( 2\pi)^3} \frac{1}{2k} \left[
(1+n_{\bf k}) \, e^{-i k (t-t')}+ n_{\bf k} \, e^{i k (t-t')}
\right]e^{i {\bf
k}\cdot ({\bf x}-{\bf x'})} \, ; \nonumber \\
&& G^{<} ({\bf x}-{\bf x'};t-t') = \int \frac{ d^3 {\bf k}}{ (
2\pi)^3} \langle \phi_{-\bf k} (t') \phi_{\bf k} (t) \rangle \,
e^{i {\bf
k}\cdot ({\bf x}-{\bf x'})} \,  \nonumber \\
&=&  \int \frac{ d^3 {\bf k}}{ ( 2\pi)^3} \frac{1}{2 k} \left[
n_{\bf k} \, e^{-i k (t-t')}+ (1+n_{\bf k}) \,  e^{i k (t-t')}
\right] e^{i {\bf k}\cdot ({\bf x}-{\bf x'})} \, .
\label{freegreen}
\end{eqnarray}
The point ${\bf \bar x}=(x,y,-z)$ is the mirror image of the point
${\bf  x}=(x,y,z)$ with respect to the unperturbed mirror's
surface $S_0$ at the $z=0$ plane.
 From Eq.(\ref{g0g1}), we can derive the following useful
identities:
\begin{eqnarray}
\partial_{t'} G_0^{< (>) }({\bf x},{\bf x'}; t,t') \mid_{z'=0}
&=& \partial_{x'\,  {\rm or} \, y'} G_0^{< (>) }({\bf x},{\bf x'};
t,t') \mid_{z'=0} = 0 \, , \nonumber \\
\partial_{z'} G_0^{< (>) }({\bf x},{\bf x'}; t,t') \mid_{z'=0}
&=&  2 \, \partial_{z'} G^{< (>) }({\bf x},{\bf x'}; t,t')
\mid_{z'=0} \, ,  \label{g0g}
\end{eqnarray}
where the Green's functions are evaluated on the mirror's surface
$S_0$. The momentum integral can be carried out to obtain the
Green's functions in terms of space and time as:
\begin{eqnarray}
G^{>} ({\bf x}-{\bf x'};t-t') &=&  {\rm Re} \left[ G ({\bf x}-{\bf
x'};t-t')\right] +i \, {\rm Im} \left[ G ({\bf x}-{\bf
x'};t-t')\right] \, , \nonumber \\
G^{<} ({\bf x}-{\bf x'};t-t') &=& {\rm Re} \left[ G ({\bf x}-{\bf
 x'};t-t') \right]-i \, {\rm Im} \left[ G ({\bf x}-{\bf x'};t-t') \right] \,
 ,\label{green}
\end{eqnarray}
where
\begin{eqnarray}
{\rm Re } \left[ G ({\bf x}-{\bf x'};t-t')\right]&=& \frac{\pi
k_{\rm B} T }{ 8 \pi^2 \mid {\bf x}-{\bf x'} \mid } \left\{\,
\coth
 \left[  \pi k_{\rm B} T \left( \, t-t'+ \mid {\bf x}-{\bf
 x'} \mid  \, \right)\right] \right. \nonumber \\
 &-&   \left. \coth
 \left[ \pi k_{\rm B} T \left(  \, t-t'-\mid {\bf x}-{\bf
 x'} \mid  \, \right)\right] \, \right\} \, , \nonumber \\
 {\rm Im } \left[ G ({\bf x}-{\bf x'};t-t')\right]&=& \frac{1 }
 { 8 \pi^2 \mid {\bf x}-{\bf x'} \mid } \left\{ \,  \delta
 \left[   \,  t-t'+ \mid {\bf x}-{\bf
 x'} \mid  \, \right] \right.  \nonumber \\
 &-&   \left. \delta
 \left[ \, t-t'- \mid {\bf x}-{\bf
 x'} \mid \, \right] \, \right\} \, . \label{reimgreen}
\end{eqnarray}

Up to order ${\cal O} (\delta q^2)$ consistent with the
approximation on the Lagrangian in Eq.(\ref{lagrangian}), the influence functional is given
by
\begin{eqnarray}
{\cal F} \left[ \delta q^+, \delta q^- \right]&=& \exp \left\{ {i
 \, \int dt \left[ \frac{1}{2} (\delta q^+ (t))^2 \langle
\frac{\partial F}{\partial q} \rangle (t)- \frac{1}{2} (\delta q^-
(t) )^2 \langle \frac{\partial F}{\partial q} \rangle  (t)
\right]}  \right. \nonumber \\
&-& \left. \frac{1}{2}  \, \int \, dt \int  \, dt' \left[
\delta q^+ (t) \, \langle F^+_0 (t) F^+_0 (t') \rangle \, \delta
q^+ (t') + \delta q^- (t) \, \langle F^-_0 (t) F^-_0 (t') \rangle
\, \delta q^- (t') \right. \right.
\nonumber \\
&-& \left. \left. \delta q^+ (t) \, \langle F^+_0 (t) F^-_0 (t')
\rangle \, \delta q^- (t')- \delta q^- (t) \, \langle F^-_0 (t)
F^+_0 (t') \rangle \, \delta q^+ (t')\right]  \right\}  \label{influence1}\, .
\end{eqnarray}
The nonequilibrium force-force correlation functions are defined as follows:
\begin{eqnarray}
\langle F^+_0 (t) F^+_0 (t') \rangle &=& \langle F_0 (t) F_0 (t')
\rangle \, \Theta (t-t') +\langle F_0 (t') F_0 (t) \rangle
\, \Theta (t'-t) \, , \nonumber \\
\langle F^-_0 (t) F^-_0 (t') \rangle &=& \langle F_0 (t) F_0 (t')
\rangle \, \Theta (t'-t) +\langle F_0 (t') F_0 (t) \rangle \,
\Theta (t-t') \, , \nonumber \\
\langle F^+_0 (t) F^-_0 (t') \rangle &=& \langle \, F_0 (t') F_0
(t) \rangle \equiv {\rm Tr} \left( {\hat\rho}_{\phi} \, F_0 (t')
F_0 (t) \right) \, ,
\nonumber \\
\langle F^-_0 (t) F^+_0 (t') \rangle &=& \langle F_0 (t) F_0 (t')
\rangle \equiv {\rm Tr} \left( {\hat \rho}_{\phi}F_0 (t) F_0 (t')
\right)\, .
\end{eqnarray}
Together with Eqs.(\ref{force1})-(\ref{forcestress}), the  force
Green's functions can be written in terms of that of the  scalar
field.

To obtain the semiclassical Langevin equation, it is more
convenient to change variables  to the average and relative
coordinates:
\begin{equation}
\delta q= \frac{1}{2} ( \delta q^+ +\delta q^- ) \, , \,\,\,\,\,
\delta r=\delta q^+ -\delta q^- \, .
\end{equation}
The coarse-grained effective action defined in Eq. (\ref{noneqaction}) with the influence
functional in Eq. (\ref{influence1})
 then becomes
\begin{eqnarray}
 S [\delta q,\delta r] &=&  \int dt \, \delta r(t) \left[-m \,
{\delta \ddot q}(t) - \frac{\delta V}{\delta q} (t)+  \,
\langle \frac{\partial F}{\partial q} \rangle  \, \delta q(t)
\right. +  \left. \,  \int dt' \, \chi_{FF} (t- t') \,  \delta
q(t') \right]  \nonumber \\
& +&  \frac{i}{2} \, \int dt \, \int dt' \, \delta r(t) \,
\sigma_{FF} (t-t') \, \delta r(t') + {\cal{O}} (\delta r^3)  \, ,
\end{eqnarray}
where
\begin{eqnarray}
\chi_{FF} (t-t') &=& i \,  \Theta( t-t') \, \langle \left[ F_0
(t), F_0 (t') \right]
\rangle \, , \label{commutator} \\
  \sigma_{FF} (t-t') &=& \frac{1}{2} \, \langle
\left\{ F_0 (t), F_0 (t') \right\} \rangle \, .
\label{anticommutator}
\end{eqnarray}
We then further introduce an auxiliary quantity $ \eta (t)$, the
noise force,   with the distribution function in terms of the
Gaussian form:
\begin{equation}
P[\eta (t)] = \exp \left\{ - \frac{1}{2}  \, \int dt \, \int
dt' \, \eta (t) \, \sigma^{-1}_{FF} (t-t') \, \eta (t') \right\}
\, . \label{noisedistri}
\end{equation}
In terms of the noise force $\eta (t) $,  the above coarse-grained
action $ S $ can be written as the field integration over $\eta (t)$ given by
\begin{equation}
\exp i S  =  \int {\cal D} \eta \, P [\eta (t)] \, \exp i S_{\rm
eff} \left[ \delta q, \delta r, \eta \right] \, ,
\end{equation}
with the effective action $ S_{\rm eff}$ :
\begin{eqnarray}
S_{\rm eff} [\delta q,\delta r, \eta ] =  \int dt \,  \delta r(t)
&& \left[-m \, \delta {\ddot  q}(t)- \frac{\delta V}{\delta q}
(t)+  \, \langle \frac{\partial F}{\partial q} \rangle \,
\delta q(t)
\right.    \nonumber \\
&&  + \left.  \int dt' \, \chi_{FF} (t- t') \, \delta
q(t')+ \eta (t) \right]+   {\cal{O}} (\delta r^3) \, .
\end{eqnarray}
The semiclassical approximation requires to extremize the
effective action $ \delta S_{\rm eff}/ \delta r$ with respect to a
particular trajectory of the mirror.  The   lowest order equation
of motion for $\delta q(t)$  where the terms beyond  order $
{\cal{O}} (\delta r^3) $ are ignored, can be obtained as follows:
\begin{equation}
m  \,\delta {\ddot q}(t)+  \frac{\delta V}{\delta q} (t)-  \,
\langle \frac{\partial F}{\partial q} \rangle \, \delta q(t)-  \int dt' \, \chi_{FF} (t- t') \, \delta q(t')= \eta (t)
\, .
\label{langevin}
\end{equation}
The noise force correlation function  is of the Gaussian form
given by Eq.(\ref{noisedistri}):
\begin{equation}
\langle \eta (t) \eta (t') \rangle =  \, \sigma_{FF} (t-t')
\, .
\end{equation}
This is a typical Langevin equation. It contains all of quantum
corrections arising from the scalar field  which are linear in
$\delta q$. The terms we ignored above involve  the coupling
between $ \delta r $ and $ \delta q $. A consistent improvement
over this semiclassical Langevin equation will involve a
perturbation expansion in these terms.

Here we would like to point out that this Langevin equation
reveals two levels of backreaction effects on the dynamics of the
mirror. They are  radiation reaction induced by the motion of the
mirror as well as backreaction dissipation arising from
fluctuations in quantum fields via a fluctuation-dissipation
theorem. Both of them are
valid in a first order expansion in the mirror's displacement.
In fact, the term for motion-induced radiation reaction is given
by the variation of the {\it mean pressure force } from the
quantum field that responds to the small displacement
of the mirror. The backreaction dissipation effect involving the
nonlocal kernel obtained from the {\it  force correlations} that reflects the general
non-Markovian nature of the
pressure forces is balanced by the force fluctuations. The kernel of the dissipative force can be obtained
from the commutator of the forces in Eq.(\ref{commutator}), and
however, the autocorrelation function for the  noise forces is
given by the anticommutator of the forces in
Eq.(\ref{anticommutator}). Thus, the balance between the effects
from dissipation and fluctuation can be encoded in
the underlying fluctuation-dissipation theorem which we can
compute explicitly in this work.
In general, when the full dynamics between the mirror and quantum
fields is considered, the above two backreaction effects have to
be treated in  a self-consistent way~\cite{HU1,HU2,HU3,HU4} where
one may find the dissipation effect via a fluctuation-dissipation
relation on the uniform accelerated particle in which radiation
reaction vanishes.

\section{ backreaction forces}

We now try to compute the backreaction forces.
 Here we mainly follow the approach developed by Ford and
Vilenkin to obtain motion-induced radiation reaction~\cite{FV}.
From Eq.(\ref{forcestress}), we have
\begin{equation}
\langle \frac{\delta F}{ \delta q} \rangle ( 0^{+},t ) \,  \delta
q (t)= \langle \frac{\delta F}{ \delta q} \rangle ( 0^{-},t ) \,
\delta q (t) \,
\end{equation}
by symmetry. Thus, the motion-induced force from one side of the
mirror needs to be computed.  Technically speaking, it is known
that the expectation values of  stress tensors and  stress tensor
correlation functions will confront   short-distance divergences
in the coincidence limit.  The method of the point-splitting will
be adopted to regularize these quantities where we take the fields
in all products of the stress tensor at different points, and the
same point limit is taken after doing renormalization. To do so,
the expectation value of the motion-induced force obtained from
Eqs.(\ref{stress}) and (\ref{forcestress}) now becomes
\begin{equation}
 \langle  \frac{ \delta F}{ \delta q} \rangle \, \delta q (t) =2
 \int_A  d^2 {\bf x}_{\parallel} \, \langle \frac{ \delta T_{zz}}{ \delta q } \rangle
 \, ({\bf x}_{\parallel}, 0^-, t ) \, \delta q(t) \,
\end{equation}
with
\begin{eqnarray}
 \langle \frac{ \delta T_{zz}}{ \delta q } \rangle  \, ({\bf
x}_{\parallel}, z, t )  \, \delta q(t)
  &=&
\frac{1}{4}(\partial_{t} \partial_{t'}+\partial_{z}\partial_{z'}
-\partial_{x}\partial_{x'}-\partial_{y}\partial_{y'})\, \nonumber
\\
 && \left[ \, \langle\phi_{0}({\bf x},t )\,\delta\phi({\bf x'},t'
)\rangle  + \langle\phi_{0}({\bf x'},t')\, \delta\phi({\bf
x},t)\rangle   \right. \nonumber \\
&&  + \left.     \langle\delta\phi({\bf x},t)\,\phi_{0}({\bf
x'},t')\rangle +  \langle\delta\phi({\bf x'},t')\,\phi_{0}({\bf
x},t)\rangle\ \right] \mid_{ {\bf x'}_{\parallel} \rightarrow {\bf
x}_{\parallel}, \, z' \rightarrow z , \, t' \rightarrow
t+\epsilon} \,, \nonumber \\
\label{stresspointsplit}
\end{eqnarray}
where $ \epsilon$ is introduced for the point-splitting method.
 The limit of $ \epsilon
\rightarrow 0$  will be taken, and the motion-induced force
expects to be finite in this limit~\cite{FV}. The perturbed field
due to the motion of the mirror  in Eq.(\ref{perturbedfield}) can
be written involving the retarded Green's function as (see
Ref.{\cite{FV}} for details):
\begin{equation}
\delta\phi({\bf x}_{\parallel}, z, t) \mid_{ z \rightarrow 0^- }=-
\int dt'\int_A d^2 {\bf x}'_{\parallel} \, \,
\partial{}_{z'}G_{0}^{\rm Ret}({\bf x},{\bf x'};t,t')
\,\partial_{z'}\phi_{0}({\bf x'},t') \, \delta q(t') \mid_{z',z
\rightarrow 0^-} \, . \label{deltaphigreen}
\end{equation}
 The retarded Green's
function is defined to be
\begin{eqnarray}
 G_{0}^{\rm Ret}({\bf x},{\bf x'} ; t,t')&& \equiv
 i\,\Theta(t-t')\langle \, [\phi_0 ({\bf x},t ),\phi_0 ({\bf x'},t')]
\, \rangle
\, \nonumber \\
&& = i\,\Theta(t-t') \left[ \, G_{0}^{>}({\bf x},{\bf x'};
t,t')-G_{0}^{<}({\bf x},{\bf x'}; t,t') \, \right]\,.
\label{green1}
\end{eqnarray}
Putting all together,
 the motion-induced force term becomes
\begin{eqnarray}
 \langle  \frac{ \delta F}{ \delta q} \rangle \, \delta q (t) &=&-
 \int^t
 dt' \delta q(t')  \left\{  8   \int_A  d^2 {\bf x}_{\parallel}  \int_A d^2
 {\bf x}'_{\parallel} \, \right.  \nonumber \\
 &&  \left[ \, \partial_{t}\partial_{z'}
 {\rm Im} \left[G ({\bf x}-{\bf x'};  t-t'+\epsilon) \right] \, \partial_{t}\partial_{z'}
 {\rm Re} \left[ G ({\bf x}-{\bf x'}; t-t') \right] \right.  \nonumber \\
 &&
 +\, \partial_{z}\partial_{z'} {\rm Im} \left[ G ({\bf x}-{\bf x'};
 t-t'+\epsilon)\right] \,
 \partial_{z}\partial_{z'}{\rm Re} \left[G ({\bf x}-{\bf x'}; t-t')
 \right]
 \nonumber \\
 &&
 -\, \partial_{x}\partial_{z'} {\rm Im} \left[ G ({\bf x}-{\bf x'};
 t-t'+\epsilon)\right] \,
 \partial_{x}\partial_{z'}{\rm Re} \left[G ({\bf x}-{\bf x'}; t-t')
 \right]
 \nonumber \\
 &&
 - \, \partial_{y}\partial_{z'} {\rm Im} \left[ G ({\bf x}-{\bf x'};
 t-t'+\epsilon)\right] \,
 \partial_{y}\partial_{z'}{\rm Re} \left[G ({\bf x}-{\bf x'}; t-t')
 \right]
 \nonumber \\
&& + \partial_{t}\partial_{z'}  {\rm Im} \left[G ({\bf x}-{\bf
x'}; t-t') \right] \, \partial_{t}\partial_{z'}
 {\rm Re} \left[ G ({\bf x}-{\bf x'}; t-t'+\epsilon) \right]  \nonumber \\
 &&
 +\, \partial_{z}\partial_{z'} {\rm Im} \left[ G ({\bf x}-{\bf x'};
 t-t')\right] \,
 \partial_{z}\partial_{z'}{\rm Re} \left[G ({\bf x}-{\bf x'}; t-t'+\epsilon)
 \right]
 \nonumber \\
 &&
 -\, \partial_{x}\partial_{z'} {\rm Im} \left[ G ({\bf x}-{\bf x'};
 t-t')\right] \,
 \partial_{x}\partial_{z'}{\rm Re} \left[G ({\bf x}-{\bf x'}; t-t'+\epsilon)
 \right]
 \nonumber \\
 && \left. \left.
 - \, \partial_{y}\partial_{z'} {\rm Im} \left[ G ({\bf x}-{\bf x'};
 t-t')\right] \,
 \partial_{y}\partial_{z'}{\rm Re} \left[G ({\bf x}-{\bf x'};
 t-t'+\epsilon)
 \right] \, \right]  \right\}\mid^{\epsilon \rightarrow 0}_{ z',z \rightarrow 0^-} \, , \label{motioninduced}
\end{eqnarray}
where we have used Eqs.(\ref{g0g}) and (\ref{green}). The
retardation effect is included as the time integration in $t'$
runs  to the time $ t$.  In addition, the force above will be
evaluated at the surface of the mirror by taking the limits of $
z', z \rightarrow 0^-$. It will suffer from short-distance
divergences that we will discuss later~\cite{FV}.

The force-force correlation function evaluated on the unperturbed
mirror's surface $S_0 $ at rest can be expressed as:
\begin{eqnarray}
\langle F_0 (t) F_0 (t') \rangle &=& \langle F_0 (t) F_0 (t')
\rangle - \langle F_0 (t) \rangle \langle F_0 (t') \rangle
\nonumber \\
&=& 2 \left[ \langle F_0 (0^-, t) F_0 (0^-, t') \rangle - \langle
F_0 (0^-, t) \rangle \langle F_0 (0^-, t') \rangle \right] \, ,
\end{eqnarray}
where we have used the fact that for a static mirror the mean
pressure force vanishes. In addition,  the force-force
correlations of each side of the mirror are the same  by symmetry,
and there is no correlation between the forces from  opposite
sides of the mirror, namely,
\begin{eqnarray}
&& \langle F_0 (0^-, t) F_0 (0^-, t') \rangle - \langle F_0 (0^-,
t) \rangle \langle F_0 (0^-, t') \rangle = \langle F_0 (0^+, t)
F_0 (0^+, t') \rangle - \langle F_0 (0^+, t) \rangle \langle F_0
(0^+,
t') \rangle \, , \nonumber \\
 && \langle F_0 (0^{\mp}, t) F_0 (0^{\pm}, t') \rangle = \langle F_0 (0^{\mp},
t) \rangle  \langle F_0 (0^{\pm}, t') \rangle \, .
\end{eqnarray}
However, one may expect that motion-induced radiation reaction  on
opposite sides of the mirror might have correlations as they both
arise due to the motion of the mirror even though the induced
forces  on opposite sides of the mirror can not communicate with
each other. This correlation effect will  contribute to the
Langevin equation where it is  of order $ {\cal{O}} (\delta q^3)
$, and can be neglected here~\cite{L}.

Using Eq.(\ref{stress0}),  we can write the above correlation
functions in terms of the Green's functions of the scalar field
given by:
\begin{eqnarray}
\langle F_0 (t) F_0 (t') \rangle &=& 2 \int_A d^2 {\bf
x}_{\parallel} \int_A d^2 {\bf x}'_{\parallel} \left[ \langle
T_{zz} ( {\bf x}_{\parallel}, 0^-; t)  T_{zz} ( {\bf
x'}_{\parallel} , 0^-; t') \rangle - \langle T_{zz} ( {\bf
x}_{\parallel}, 0^-; t) \rangle \langle T_{zz} ( {\bf
x'}_{\parallel}, 0^-; t') \rangle
\right] \, , \nonumber \\
 \langle F_0 (t') F_0 (t) \rangle &=& 2  \int_A
d^2 {\bf x}_{\parallel}  \int_A d^2 {\bf x}'_{\parallel} \left[
\langle T_{zz} ( {\bf x'}_{\parallel}, 0^-; t')  T_{zz} ( {\bf
x}_{\parallel} , 0^-; t) \rangle - \langle T_{zz} ( {\bf
x'}_{\parallel}, 0^-; t') \rangle \langle T_{zz} (  {\bf
x}_{\parallel}, 0^-; t) \rangle
\right] \, , \nonumber \\
\end{eqnarray}
where
\begin{eqnarray}
&& \langle T_{zz} ( {\bf x}_{\parallel}, z;  t)  T_{zz} ({\bf
x'}_{\parallel}, z'; t') \rangle - \langle T_{zz} ( {\bf
x}_{\parallel}, z; t) \rangle \langle T_{zz} ( {\bf
x'}_{\parallel} , z'; t') \rangle  \nonumber \\
&&=  \frac{1}{4} \left\{ (\partial_t
\partial_{t''} + \partial_z \partial_{z''} -\partial_x
\partial_{x''} -\partial_y
 \partial_{y''})  (\partial_{t'} \partial_{t'''} + \partial_{z'}
\partial_{z'''} -\partial_{x'} \partial_{x'''} -\partial_{y'}
\partial_{y'''}) \right. \nonumber \\
&& \times \left. \left[ \, G_0^{>} ({\bf x},{\bf x'}; t,t')\,
G_0^{>} ({\bf x''},{\bf x'''}; t'',t''' ) + G_0^{>} ({\bf x},{\bf
x'''}; t,t''') \, G_0^{>}({\bf x''},{\bf x'}; t'',t')\, \right]
\right\} \mid^{{\bf{x}}''_{\parallel} \rightarrow {\bf{x}}
_{\parallel},z'' \rightarrow z, t'' \rightarrow t+ \epsilon'
}_{{\bf{x}}'''_{\parallel} \rightarrow {\bf{x}}'_
{\parallel}, z''' \rightarrow z', t''' \rightarrow t'+\epsilon''} \, ,\nonumber \\
&& \langle T_{zz} ( {\bf x'}_{\parallel}, z'; t')  T_{zz} ({\bf
x}_{\parallel}, z ; t) \rangle - \langle T_{zz} ( {\bf
x'}_{\parallel}, z'; t') \rangle \langle T_{zz} ( {\bf
x}_{\parallel} , z; t) \rangle  \nonumber \\
&&= \frac{1}{4} \left\{ (\partial_t
\partial_{t''} + \partial_z \partial_{z''} -\partial_x
\partial_{x''} -\partial_y
 \partial_{y''})  (\partial_{t'} \partial_{t'''} + \partial_{z'}
\partial_{z'''} -\partial_{x'} \partial_{x'''} -\partial_{y'}
\partial_{y'''}) \right. \nonumber \\
&& \times \left. \left[ \, G_0^{<} ({{\bf x}},{\bf x'}; t,t')\,
G_0^{<} ({\bf x''},{\bf x'''}; t'',t''' ) + G_0^{<} ({\bf x},{\bf
x'''}; t,t''') \, G_0^{<}({\bf x''},{\bf x'}; t'',t') \, \right]
\right\} \mid^{{\bf{x}}''_{\parallel} \rightarrow {\bf{x}}
_{\parallel},z'' \rightarrow z, t'' \rightarrow t+ \epsilon'
}_{{\bf{x}}'''_{\parallel} \rightarrow {\bf {x}}'_ {\parallel},
z'''
\rightarrow z', t''' \rightarrow t'+\epsilon''   } \, . \nonumber \\
\end{eqnarray}
The limits  of $ \epsilon', \epsilon'' \rightarrow 0 $ due to the
point-splitting  will be taken. We then evaluate the correlation
functions  at the surface of the mirror by taking the limits of  $
z,z' \rightarrow 0^-$.  The above expressions can be simplified
with Eq.(\ref{g0g}) as
\begin{widetext}
\begin{eqnarray}
\langle F_0 (t) F_0 (t') \rangle &=& 2 \int_A d^2 {\bf
x}_{\parallel} \int_A d^2 {\bf x}'_{\parallel}  \left[ \,
\partial_t
\partial_{z'}  G^> ({\bf x}-{\bf x'}; t-t') \,\,
\partial_{t} \partial_{z'} G^> ({\bf x}-{\bf x'};
t-t'+ (\epsilon'-\epsilon'')) \right. \nonumber \\
&& + \partial_z
\partial_{z'} G^> ({\bf x}-{\bf x'}; t-t') \,\,
\partial_{z} \partial_{z'} G^> ({\bf
x}-{\bf x'}; t-t'+(\epsilon'-\epsilon''))  \nonumber \\
&& -\partial_x \partial_{z'} G^> ({\bf x}-{\bf x'}; t-t') \,\,
\partial_{x} \partial_{z'} G^> ({\bf x}-{\bf x'};
t-t'+ (\epsilon'-\epsilon'')) \nonumber \\
&& - \partial_y \partial_{z'} G^> ({\bf x}-{\bf x'}; t-t') \,\,
\partial_{y} \partial_{z'} G^> ({\bf
x}-{\bf x'}; t-t'+(\epsilon'-\epsilon'')) \nonumber \\
&& +  \partial_t
\partial_{z'}  G^> ({\bf x}-{\bf x'}; t-t'-\epsilon'') \,\,
\partial_{t} \partial_{z'} G^> ({\bf x}-{\bf x'};
t-t'+ \epsilon' )  \nonumber \\
&& + \partial_z
\partial_{z'} G^> ({\bf x}-{\bf x'}; t-t'-\epsilon'') \,\,
\partial_{z} \partial_{z'} G^> ({\bf
x}-{\bf x'}; t-t'+\epsilon')  \nonumber \\
&& -\partial_x \partial_{z'} G^> ({\bf x}-{\bf x'};
t-t'-\epsilon'') \,\,
\partial_{x} \partial_{z'} G^> ({\bf x}-{\bf x'};
t-t'+ \epsilon') \nonumber \\
&& \left.  - \partial_y \partial_{z'} G^> ({\bf x}-{\bf x'};
t-t'-\epsilon'') \,\,
\partial_{y} \partial_{z'} G^> ({\bf
x}-{\bf x'}; t-t'+ \epsilon') \, \right]
\mid^{\epsilon', \epsilon'' \rightarrow 0}_{ z', z'' \rightarrow 0} \, ,  \nonumber \\
\langle F_0 (t') F_0 (t) \rangle &=& 2 \int_A  d^2 {\bf
x}_{\parallel} \int_A  d^2 {\bf x}'_{\parallel}  \left[ \,
\partial_t
\partial_{z'}  G^< ({\bf x}-{\bf x'}; t-t') \,\,
\partial_{t} \partial_{z'} G^< ({\bf x}-{\bf x'};
t-t'+ (\epsilon'-\epsilon'')) \right. \nonumber \\
&& + \partial_z
\partial_{z'} G^< ({\bf x}-{\bf x'}; t-t') \,\,
\partial_{z} \partial_{z'} G^> ({\bf
x}-{\bf x'}; t-t'+(\epsilon'-\epsilon''))  \nonumber \\
&& -\partial_x \partial_{z'} G^< ({\bf x}-{\bf x'}; t-t') \,\,
\partial_{x} \partial_{z'} G^< ({\bf x}-{\bf x'};
t-t'+ (\epsilon'-\epsilon'')) \nonumber \\
&& - \partial_y \partial_{z'} G^< ({\bf x}-{\bf x'}; t-t') \,\,
\partial_{y} \partial_{z'} G^< ({\bf
x}-{\bf x'}; t-t'+(\epsilon'-\epsilon'')) \nonumber \\
&& +  \partial_t
\partial_{z'}  G^< ({\bf x}-{\bf x'}; t-t'-\epsilon'') \,\,
\partial_{t} \partial_{z'} G^< ({\bf x}-{\bf x'};
t-t'+ \epsilon' )  \nonumber \\
&& + \partial_z
\partial_{z'} G^< ({\bf x}-{\bf x'}; t-t'-\epsilon'') \,\,
\partial_{z} \partial_{z'} G^< ({\bf
x}-{\bf x'}; t-t'+\epsilon')  \nonumber \\
&& -\partial_x \partial_{z'} G^< ({\bf x}-{\bf x'};
t-t'-\epsilon'') \,\,
\partial_{x} \partial_{z'} G^< ({\bf x}-{\bf x'};
t-t'+ \epsilon') \nonumber \\
&& \left.  - \partial_y \partial_{z'} G^< ({\bf x}-{\bf x'};
t-t'-\epsilon'') \,\,
\partial_{y} \partial_{z'} G^< ({\bf
x}-{\bf x'}; t-t'+ \epsilon') \, \right] \mid^{\epsilon',
\epsilon'' \rightarrow 0}_{ z',  z \rightarrow 0} \, ,
\label{ffcorrelation}
\end{eqnarray}
\end{widetext}
In this stage, we can compute the commutator, $ \chi_{FF}$  and
the anticommutator, $ \sigma_{FF}$ of the forces  in
Eqs.(\ref{commutator}) and (\ref{anticommutator}) respectively
which allow us to discuss the issue of the fluctuation-dissipation
theorem below. In particular, it is a straightforward calculation
to show that
\begin{equation}
\langle  \frac{ \delta F}{ \delta q} \rangle \delta q (t) = \int
dt' \, \chi_{FF} ( t-t') \, \delta q(t') \, ,
\label{motioninduced-kernel}
\end{equation}
using the Green's functions of the scalar field in
Eqs.(\ref{green}) and (\ref{reimgreen}). The above relation holds
only for the mirror with the small displacement where the coupling of the mirror to
the quantum field is quadratic in field variables. Similar
result that relates these two backreaction effects has been found in Ref \cite{JR2}
 in the 1+1 dimensional spacetime.
 Even though the fluctuation-dissipation relation we will discuss later can link the
 dissipation effect obtained from   force correlations to the force
 fluctuations, one cannot conclude that radiation reaction due to the motion of
the mirror is balanced by the force fluctuations. Notice
that it has been recently mentioned by Hu~\cite{HU1,HU2,HU3} that
there are incorrect claims in which radiation reaction is balanced
by the force fluctuations. We would like to  emphasize
that Eq.(\ref{motioninduced-kernel}) does not hold for the general
situations of  couplings. For example, one can consider the
coupling between the mirror and the quantum field which is
proportional to  field variables to  the $n$-th power. For an odd number of the power,
it is obvious that diagrammatically the above
radiation reaction vanishes as the dissipation effect
from  force fluctuations gives  the non-zero
contribution to the Langevin equation. However, as for an even
number of the  power, since the effect of
radiation reaction is given by the $(n/2)$-loop integral while the dissipation effect
is given by the $(n-1) $-loop integral, two backreaction effects can possibly be
equal only for $n=2$.

\section{ Fluctuation-dissipation theorem}

Fluctuation-dissipation theorem plays a vital role in balancing
between these two effects to dynamically stabilize a
nonequilibrium Brownian motion in the presence of external
fluctuation forces. In the case of  classical Brownian motion, the
nonequilibrium dynamics of the Brownian object moving in a
stationary fluid can be described by a phenomenological Langevin
equation. Incessant collisions from the molecules of the fluid
with the Brownian object produce both resistance to the motion of
the object and fluctuations in its trajectory. The Langevin
equation can account for these two effects by introducing friction
and dissipation as well as a stochastic force as below:
\begin{equation}
\ddot{q} (t) + \gamma  \,  \dot{q} (t) = \eta (t) \, ,
\end{equation}
where the dissipative force is given by the  time derivative of
the position  with the damping coefficient $\gamma$, and $ \eta
(t)$ stands for a stochastic   force that mimics  random kicks of
the  molecules on the Brownian object with  white noise
properties:
\begin{eqnarray}
\langle \eta (t) \rangle  &=& 0 \, ; \nonumber \\
\langle \eta(t) \eta (t') \rangle &=& 2 \gamma k_{ \rm B} T \,
\delta (t-t')\, .
\end{eqnarray}
$ k_{\rm B}$ is Boltzmann constant and the average is taken with
respect to the thermal ensemble of  fluctuations of the fluid at
temperature $T$. The dissipation and fluctuation kernels can be
defined respectively as:
\begin{eqnarray}
&& \gamma \, \dot{q}  (t) = - \int dt'  \mu ( t-t') q(t') \, \, ,
\,\,\,\,\,\,\,\, \mu (t-t') = - \gamma \frac{d}{d t} \delta (t-t')
\, ;
\nonumber \\
&& \langle \eta (t) \eta (t') \rangle = \nu (t-t') \, \,
,\,\,\,\,\,\,\,\,
  \nu (t-t')= 2 \gamma k_{ \rm B} T \delta (t-t') \, .
\end{eqnarray}
The fluctuation-dissipation theorem is to relate the dissipation
kernel to the fluctuation kernel of the form :
\begin{equation}
\mu (t-t') =- \frac{1}{ 2  k_{ \rm B} T} \frac{d}{d t} \nu (t-t')
\, , \label{df-classical}
 \end{equation}
 which is independent of the spectrum density of  thermal fluid
 and the coupling strength of the Brownian object with the molecules
 in a fluid.

A very clear microscopic description  to the Langevin equation
within the context of one-particle quantum mechanics coupled to a
bath of harmonic oscillators has been presented by Caldeira and
Leggett~\cite{CL}. Using the Feynman-Vernon influence functional,
their study reveals that in general the dissipation term arises
from a local approximation to the non-Markovian kernel for a
particular choice of the density of states of the heat bath, and
as a result,  the noise forces  become uncorrelated over
macroscopic time scales larger than the typical scales determined
by the bath. They are thus related by the above classical
fluctuation and dissipation theorem. Recently, the studies have
been devoted to this issue where the Brownian object is coupled to
quantum fields. The coupling between the Brownian object and
quantum fields  is assumed to be linear or nonlinear in terms of
the variable of the Brownian object, and it is linear in terms of
the field variable in which the field in momentum space can be
treated as a bath of harmonic
oscillators~\cite{HU1,HU2,HU3,HU4,JR1}. However, the case we
consider is more complicated since the coupling of the mirror to
quantum fields is given by the area-integral of the  stress tensor
which is quadratic in fields~\cite{JR2,MMF}. In the presence of a
perfectly reflecting mirror, we impose an idealized boundary
condition on quantum fields where the fields  vanish on the
surface of the mirror. This unrealistic boundary condition  in
fact leads to a troublesome result: the  stress tensor is
divergent when it is evaluated on  the surface of the
mirror~\cite{FV,WF}. Hence we need to introduce a cut-off on $z$,
the distance to the mirror, thus resulting in some complications
as we try to derive the corresponding  fluctuation-dissipation
theorem from a microscopic point of view.

To obtain the corresponding  fluctuation and dissipation theorem
, we take the Fourier transform of $ \chi_{FF}
(t-t') $ and $ \sigma_{FF} (t-t')$ as:
 \begin{eqnarray}
 \chi_{FF} (t-t') &=& \int \frac{d w}{2\pi} \, \chi_{FF} ( \omega) \, e^{ -i
 \omega (t-t') } \, , \nonumber \\
 \sigma_{FF} (t-t') &=& \int \frac{d w}{2\pi} \, \sigma_{FF} ( \omega) \, e^{ -i
 \omega (t-t') } \, . \label{ft}
\end{eqnarray}
 Then, we introduce the spectral density $ \rho (\omega)$ of quantum field
 defined to be:
 \begin{equation}
 \chi_{FF} ( \omega) = \int \frac{d w'}{2\pi} \frac{ \rho (
 \omega')}{ w-w'+i \delta}  \, , \label{xirho}
 \end{equation}
 where
 the $  i \delta $ prescription is introduced to account for the
 retardation effect as the limit of $ \delta \rightarrow 0^+ $ is taken.
Thus,  substituting Eq.(\ref{xirho}) into Eq.(\ref{ft}) leads to
\begin{equation}
\chi_{FF} (t-t') =-i \, \Theta (t-t') \int \frac{d \omega}{ 2 \pi}
\, \rho (\omega) \, e^{-i \omega (t-t') } \, . \label{xirho1}
 \end{equation}
The Fourier transform of the Green's functions ( $ G^{>} , G^{<} $
) in Eq.(\ref{freegreen})
 are given by
\begin{equation}
G^{ > (<)}  ( {\bf x}-{\bf x'}; t-t') = \int \frac{d w}{2\pi} \int
\frac{d^3 {\bf k}} { (2\pi)^3} \, g^{> (<)} ( {\bf k}, \omega) \,
e^{ -i \omega (t-t') } \, e^{ i {\bf k} \cdot ({\bf x}-{\bf x'}) }
\, ,
\end{equation}
where
\begin{eqnarray}
g^> ( {\bf k}, \omega ) &=& \frac{1}{ 2 k} \, [ \, ( 1+ n_k ) \,
\delta (
\omega-k) + n_k \, \delta ( \omega +k) \, ] \, , \nonumber \\
g^< ( {\bf k}, \omega ) &=& \frac{1}{ 2 k} \, [ \, ( 1+ n_k ) \,
\delta ( \omega+ k) + n_k \, \delta ( \omega -k)  \, ] \, ,
\label{g>g<}
\end{eqnarray}
with $ k= \mid {\bf k} \mid $, and $n_k= ( e^{\beta k }-1 )^{-1}
,$ the Bose-Einstein distribution function. Then, they obey the
KMS relation~\cite{BE} given by
\begin{equation}
g^< ( {\bf k}, \omega ) = e^{ - \beta \omega} g^> ( {\bf k},
\omega ) \, . \label{KMS}
\end{equation}
Using Eq.(\ref{xirho1}), the spectral density can be obtained from
$ \chi_{FF} (t-t')$  given by the commutator of the forces from
Eqs.(\ref{commutator}) and (\ref{ffcorrelation}) as:
\begin{eqnarray}
\rho (\omega) &=& - 2 A \int \frac{ d^2 {\bf k}_{\parallel}}{ (
2\pi)^2} \int \frac{d k_{\perp}}{ 2\pi} \frac{d k'_{\perp}}{ 2\pi}
\int \frac{ d \omega'}{ 2 \pi} \, \left[ \, k_{\perp} k'_{\perp} (
\, \omega' (
\omega-\omega') + k_{\perp} k'_{\perp} + {\bf k}_{\parallel}^2 \, ) \,  \right] \nonumber \\
&& \times \, g^> ( {\bf k}_{\parallel}, k_{\perp}, \omega-\omega')
\, g^> (
-{\bf k}_{\parallel}, k'_{\perp}, \omega')   \left[ 1- e^{-\beta \omega} \right] \nonumber \\
&& \times \, e^{i ( k_{\perp}+ k'_{\perp}) ( z-z')} \, e^{-i
\omega' ( \epsilon'-\epsilon'')} \left( 1+ e^{ i \omega
\epsilon''} \right) \mid^{ \epsilon', \epsilon'' \rightarrow
0}_{z' , z \rightarrow 0^-} \, .
\end{eqnarray}
The fluctuation-dissipation theorem can be  obtained before
taking the short-distance limits. It is then a straightforward
calculation to obtain the relation between the Fourier transform
of the anticommutator of the forces $\sigma_{FF} (\omega) $ from
Eqs.(\ref{anticommutator}) and (\ref{ffcorrelation}) and the
spectral density $ \rho ( \omega) $ above. The fluctuation and
dissipation
 theorem is to link the Fourier transform of the fluctuation
kernel, the anticommutator of the forces $\sigma_{FF} (\omega) $,
to the imaginary part of the dissipation kernel, the commutator of
the forces $ \chi_{FF} ( \omega ) $,  as follows:
\begin{equation}
\sigma_{FF} ( \omega ) = - \frac{1}{2} \rho ( \omega)  \coth
\left[ \frac{\beta \omega}{ 2} \right] = {\rm Im} \left[ \chi_{FF}
( \omega) \right] \coth \left[ \frac{ \beta \omega}{2} \right] \,
\label{df-T} .
\end{equation}
The above relation relies on the fact that
\begin{equation}
 {\rm Im} \left[ \chi_{FF}
( \omega) \right] = - \frac{1}{2} \rho ( \omega) \, ,
\end{equation}
as a result of Eq.(\ref{xirho}). The high-T limit can be taken and
then the   fluctuation and dissipation theorem in this limit
reduces to
\begin{equation}
 {\rm Im} \left[ \chi_{FF}
( \omega) \right] = \frac{\omega} { 2 k_{\rm B} T} \sigma_{FF} (
\omega ) \, . \label{df-high-T}
\end{equation}
As expected,  it corresponds to the  classical Brownian motion
which can be seen by taking the Fourier transform of
Eq.(\ref{df-classical}).

The fluctuation and dissipation effects driven by quantum fields
in vacuum on a microscopic object are of great interest in regard
to imposing fundamental limits on the uncertainty of the position
and velocity of an object. In vacuum, the Fourier transforms of
the Green's functions $ ( G^{>} , G^{<}) $ are found to satisfy
the following relation:
\begin{equation}
g^< ( {\bf k}, \omega ) =  g^> ( {\bf k}, -\omega ) \, ,
\end{equation}
 from taking the limit of $ T \rightarrow 0 $ in Eq.(\ref{g>g<}).
Then, it leads to the fluctuation and dissipation theorem in
vacuum given by
\begin{equation}
\sigma_{FF} ( \omega ) = - \frac{1}{2} \rho ( \omega)  \left[
\Theta (\omega) - \Theta (- \omega) \right] = {\rm Im} \left[
\chi_{FF} ( \omega) \right]  \left[ \Theta (\omega) - \Theta (
-\omega) \right] \, . \label{df-vac}
\end{equation}
This result can also be obtained by taking the limit of $ T
\rightarrow 0$ directly from Eq.(\ref{df-T}).

Although it is generally expected that the theorem of fluctuation
and dissipation is of model
independence for the case with the small mirror's displacement in the vacuum and/or
thermal states of the field such that this theorem has been used to study the
dynamics of moving mirrors in quantum fields on various situations
of couplings~\cite{JR2}, it is still worth noticing that the study
from the first principles derivation reveals  that the obtained
theorem is also {\it independent} of the short-distance regulators
introduced to deal with divergences from quantum fields. The
theorem relates these two effects in
vacuum and/or in a thermal bath regardless of the details of
short-distance divergences associated with the underlying
microscopic dynamics. Thus, when the method of regularization is
introduced to compute the dissipation and fluctuation effects,
this theorem must be fulfilled as the results are obtained by
taking the short-distance limit in the end of calculations. It
seems to play a role as the Ward identity derived from underlying
symmetry in quantum field theory where the introduction of
regularization and renormalization  to deal with divergences must
respect this identity. This theorem also allows us to compute the
dissipation kernel from the obtained fluctuation kernel  and vice
versa which we will adopt to obtain the  Langevin equation later.

\section{moving mirrors dynamics}

We are now to study the dynamics of moving mirrors in quantum
fields driven by either vacuum or thermal fluctuations
respectively. We will take advantage of the
fluctuation-dissipation theorem derived above to obtain the
Langevin equation and to solve it consistently. The same Langevin
equation can be obtained by computing the effects of fluctuation
and dissipation separately where the corresponding
fluctuation-dissipation theorem must be fulfilled as the
short-distance limit is taken.

\subsection{Vacuum fluctuations}
We compute  the dissipation kernel  from Eqs.(\ref{motioninduced})
and (\ref{motioninduced-kernel}). To do so, the Green's functions
of the scalar field in the limit of $ T \rightarrow 0$ are
obtained from Eq.(\ref{reimgreen}) as follows:
\begin{eqnarray}
{\rm Re } \left[ G ({\bf x}-{\bf x'};t-t')\right]&=& \frac{-1}{ 4
\pi^2 [ \, (\, t-t' \,)^2 -  \mid {\bf x}-{\bf x}'\mid^2 \, ]} \, , \nonumber \\
 {\rm Im } \left[ G ({\bf x}-{\bf x'};t-t')\right]&=& \frac{-1 }
 { 8 \pi^2 \mid {\bf x}-{\bf x'} \mid } \left\{ \,   \delta
 \left[ \, t-t'- \mid {\bf x}-{\bf
 x'} \mid \, \right] \, \right\} \, ,
\end{eqnarray}
where  $ {\rm Im} \left[ G
 ({\bf x}-{\bf x'}, t-t') \right] $ has included the retardation
 effect.
 The area integration over $ {\bf x}'_{\parallel} $ in
 Eq.(\ref{motioninduced}) gives the factor $ A$, area of the mirror. Taking advantage of the $\delta$-function in $
{\rm Im} \left[ G
 ({\bf x}-{\bf x'}, t-t') \right] $  allows us to carry out the area integral on $ {\bf x}_{\parallel}$  where we
assume
 that the mirror is of a disk.
 Then, after a lengthy calculation,  the dissipative force term ends up with
\begin{eqnarray}
&& \int dt' \chi_{FF} (t-t') \, \delta q(t')  =
\frac{A}{480\pi^{2}} \, \int^{t-z}_{0} dt' \, \left[ \,
\frac{75\,\delta q(t')}{(t-t')^{6}} +\frac{75 \,\delta
q'(t')}{(t-t')^{5}} +\frac{30\,\delta q''(t')}{(t-t')^{4}}
+\frac{5\,\delta q^{[3]}(t')}{(t-t')^{3}} \right. \nonumber \\
&& -z^2 \left( \, \frac{1575\,\delta q(t')}{(t-t')^{8}}
+\frac{1575\,\delta q'(t')}{(t-t')^{7}} +\frac{675\,\delta
q''(t')}{(t-t')^{6}} +\frac{150\,\delta q^{[3]}(t')}{(t-t')^{5}}
+ \frac{15\,\delta q^{[4]}(t')}{(t-t')^{4}} \, \right) \nonumber \\
&&  \left. +z^4 \left( \, \frac{1890\,\delta q(t')}{(t-t')^{10}}
+\frac{1890\,\delta q'(t')}{(t-t')^{9}} +\frac{840\,\delta
q''(t')}{(t-t')^{8}} +\frac{210\,\delta q^{[3]}(t')}{(t-t')^{7}}
+\frac{30\,\delta q^{[4]}(t')}{(t-t')^{6}}
+\frac{2\,\delta q^{[5]}(t')}{(t-t')^{5}} \, \right) \right] \,, \nonumber \\
&&\label{eq:delta_T_t}
\end{eqnarray}
where the limit of $\epsilon \rightarrow 0 $ has been taken. We
now perform the remaining time integral and use the relation
\begin{equation}
\int^{t+{\frac{1}{\Lambda}}}_{0} dt' \, (t-t')^{-n}\delta
q^{(m)}(t')=\frac{ (-\Lambda)^{n-1} \, \delta
q^{(m)}(t+{\frac{1}{\Lambda})}}{{(n-1)}}
-\frac{1}{n-1}\int^{t+{\frac{1}{\Lambda}}}_{0}\frac{dt'}{(t-t')^{n-1}}\delta
q^{(m+1)}(t') \,
\end{equation}
by dropping out the terms evaluated at an initial time which is
equivalent to introducing an adiabatical switch-on  interaction.
Apparently, the force cannot be evaluated infinitesimally close to
the surface of the mirror  by taking the limit of $ z\rightarrow
0^-$ due to short-distance divergences. This is mainly due to an
unrealistic perfectly reflecting  condition imposed on the mirror.
It can be solved by introducing either  a fluctuating boundary in
3+1 dimensions~\cite{FS} or a non-perfectly reflecting boundary in
1+1 dimensions~\cite{JR1,JR2}. The latter condition seems to be
not sufficient to solve the divergence problem in 3+1
dimensions~\cite{SF}. The introduced energy cutoff $\Lambda$ is to
set a cutoff on $z \approx 1/\Lambda$ due to fluctuations of the
mirror's surface. Then, a local approximation can be made as the
time scales we consider are such that $ t>> 1/\Lambda $. In
vacuum, the local dissipative force  can be obtained as:
\begin{equation}
\int dt' \chi_{FF} (t-t') \, \delta q(t')  =\frac{A}{48 \pi^2}
\left(  \Lambda^3  \,\delta \ddot{q} (t) - \frac{\Lambda}{10} \,
\delta q^{[4]} (t)  - \frac{1}{15} \, \delta q^{[5]} (t)  +
{\cal{O}} \left( \frac{1}{\Lambda} \right)   \right) \, .
\end{equation}
Then, the dissipation kernel can be read off as:
\begin{equation}
\chi_{FF} (t-t') =   \frac{A}{48 \pi^2} \left(  \Lambda^3 \,
\delta^{[2]} (t-t') - \frac{\Lambda}{10} \, \delta^{[4]} (t-t') -
\frac{1}{15} \, \delta^{[5]} (t-t')    \right) \, ,
\end{equation}
where the derivatives of the $\delta$-function are involved, and
the terms of order $ {\cal{O}} ( 1/\Lambda) $ are ignored. Using
Eq.(\ref{df-vac}), the fluctuation-dissipation theorem in vacuum,
we can obtain $ \sigma_{FF}(t-t')$ by taking the Fourier transform
of $ \sigma_{FF} ( \omega)$ as:
\begin{eqnarray}
\sigma_{FF} (t-t') &=& \int \frac{d \omega}{ 2 \pi} \, \, {\rm Im}
\left[ \chi_{FF} (\omega) \right] \,  \left[ \Theta (\omega) -
\Theta ( -\omega)
\right] \,  e^ {-i \omega ( t-t')} \, \nonumber \\
&=& \frac{A}{720 \pi^2} \int \frac{d \omega}{ 2 \pi} \omega^5 \,
\cos \left [ \omega ( t-t') \right] \, . \label{ff-vac}
\end{eqnarray}
The backreaction dissipation effect above is related to the force
fluctuations via a fluctuation-dissipation relation as in the case
of Brownian motion. In addition,  motion-induced radiation
reaction due to  nonuniform acceleration of the moving mirror can
be obtained from Eq.(\ref{motioninduced-kernel}) consistent with
the result from Ref.\cite{FV}.

Notice that the known problems of the runaway solution and
preacceleration  are in the Lorentz-Dirac theory of radiation
reaction on the motion of point charges in quantum electromagnetic
fields.  The motion-induced radiation reaction force is given by
the third time derivative of the  position, and is  to accelerate
point charges~\cite{MS,CT,FLO}. The recent studies in
Refs~\cite{HU1,HU2,HU3,HU4} have found that the non-Markovian
nature of the dissipation kernel from quantum fields plays a key
role to obtain the causal equations with free of runaway solutions
within a context of the fully  nonequilibrium open system
dynamics. However, in the case with the small mirror's displacement, as we will see,
the obtained Langevin equation below even including the Markovian
backreaction force terms  of the higher derivatives (e.g. $ \delta
q^{[n]}, n>2 $) can be solved consistently with the ordinary
Newtonian  initial data.

Then, the corresponding Langevin equation  including all
backreaction effects becomes:
\begin{equation}
m   \, \delta \ddot{q} (t) + \frac{\delta V}{\delta q} (t)+ \left[
\frac{A}{24 \pi^2} \left( - \Lambda^3  \,\delta \ddot{q} (t) +
\frac{\Lambda}{10} \, \delta q^{[4]} (t) +\frac{1}{15} \, \delta
q^{[5]} (t) \right) \right]= \eta (t) \, , \nonumber
\\
\end{equation}
with the Gaussian force correlations given by Eq.(\ref{ff-vac})
as:
\begin{equation}
 \langle \eta (t) \eta (t') \rangle = \frac{A}{720 \pi^2} \int \frac{d \omega}{ 2 \pi} \omega^5 \,
 \cos
\left [ \omega ( t-t') \right] \, . \label{ff-vac-1}
\end{equation}
 The first two terms of the backreaction effects in the Langevin equation
 will modify the dispersive part of the mirror
while the third term is a dominant dissipative force term to slow
down the motion of the mirror. In fact, the first term above can
be absorbed into the renormalization of mass given by:
\begin{equation}
m_{\rm R} = m-\frac{A}{24 \pi^2} \left( \frac{\Lambda}{ \hbar c}
\right)^2 \frac{\Lambda}{c^2} \, , \label{massrenormalized}
\end{equation}
where the energy cutoff $\Lambda$ is chosen  for having positive
renormaized mass so as to avoid the runaway solution. The
renormalized mass is a parameter here to be determined from
experiment.

We now try to solve the equation by first of all, taking the
average of the above equation to understand its relaxational
dynamics. Consider the case where the mirror is attached to a
spring and undergoes oscillations with a natural frequency $
\omega_0$. Then, the equation  can be written as:
\begin{equation}
m  \,  \delta \ddot{q} (t) + m  \omega_0^2 \, \delta q+  \left[
\frac{A}{24 \pi^2} \left( \frac{\Lambda}{10} \, \delta q^{[4]} (t)
+\frac{1}{15} \, \delta q^{[5]} (t) \right) \right]= 0 \, .
\label{meanlangevin}
\end{equation}
To see the quantum effects from the scalar field on the dynamic of the mirror driven by
the classical external potential, we write the
solution of the equation as:
\begin{equation}
\delta q(t) =\delta q_c (t) + \delta q_{\hbar} (t) \, ,
\end{equation}
where $ \delta q_c (t) $ is a solution of the equation for
harmonic oscillations, and $\delta q_{\hbar} (t)$ is derivation from its
classical trajectory induced from vacuum fluctuations due to the
presence of forth and fifth time derivatives of $\delta q_c (t)$.
 Thus, they obey the following equations respectively:
\begin{eqnarray}
m   \, \delta \ddot{q}_c  (t) + m  \omega_0^2 \, \delta q_c (t)
&=& 0 \, ,
\nonumber\\
m  \, \delta \ddot{q}_{\hbar} (t) + m  \omega_0^2 \, \delta q_{\hbar} (t)&=& -
\left[ \frac{A}{24 \pi^2} \left( \frac{\Lambda}{10} \, \delta q_{c}^{
[4]} (t) +\frac{1}{15} \, \delta q_{c}^{ [5]} (t) \right) \right] \,
.
 \label{mirror-vac}
\end{eqnarray}
The equations can be solved iteratively in terms of  the retarded
Green's function:
\begin{equation}
G_{\rm ret} (t-t') = \Theta (t-t') \left[ \frac{1}{\omega_0} \sin
\left[ \omega_0 (t-t') \right] \right] \, .
\end{equation}
Thus, the solution to Eq.(\ref{mirror-vac}) is given by
\begin{eqnarray}
\delta q_c (t) &=& l_0  \cos \left[ \omega_0 (t-\theta_0 ) \right]
\, ,
\nonumber \\
\delta q_{\hbar} (t) &=&  -  \frac{A}{24 \pi^2 m \omega_0}
\int_{t_0}^t d t' \, \sin \left[ \omega_0 (t-t') \right] \,\left(
\frac{\Lambda}{10} \, \delta q_{c}^{ [4]} (t') +\frac{1}{15} \,
\delta q_{c}^{ [5]} (t') \right) \nonumber \\
&=& l_0 \left[ -\frac{A  }{ 720 \pi^2  m } \, \omega_0^4 \,
(t-t_0)   \cos \left[ \omega_0 (t-\theta_0 ) \right]  -\frac{A }{
240 \pi^2  m } \, \omega_0^3 \Lambda \, (t-t_0) \sin \left[
\omega_0 (t-\theta_0 )\right] \right]
\nonumber \\
&+& {\rm non-secular ~terms} \, .
\end{eqnarray}
Then the  $\delta q (t) $ is obtained as:
\begin{eqnarray}
\delta q (t) &=& l_0 \left\{ \left[ 1-\frac{A  }{ 720 \pi^2  m }
\, \omega_0^4 \, (t-t_0) \right]  \cos \left[ \omega_0 (t-\theta_0
) \right] -\frac{A  }{ 240 \pi^2  m } \, \omega_0^3 \Lambda \,
(t-t_0) \sin \left[ \omega_0 (t-\theta_0 )\right] \right\}
\nonumber \\
&+& {\rm non-secular ~terms} \, . \label{baresol}
\end{eqnarray}
The initial time is set at $ t_0$ and the parameters, $ l_0$ and
$\theta_0$,  can be determined by the initial conditions. Note
that the naive perturbation contains the secular terms that grow
linearly in time while the terms denoted by non-secular terms are
finite at all times. It indicates that the perturbation breaks
down at late times. In order to obtain the solution with the
correct damping behavior,  the method  of dynamical
renormalization group will be invoked to resum these secular terms
consistently~\cite{BD2}. The dynamical renormalization can be
achieved by introducing an arbitrary  time scale $ \tau$ ,
splitting $ t-t_0$ as $t-\tau+\tau-t_0$, and absorbing the terms
containing $ \tau-t_0$ into  renormaization of the amplitude $ l
(\tau) $ and the phase $ \theta (\tau) $ respectively. We then
relate $ l_0$ and $ \theta_0 $ to $ l(\tau)$ and $ \theta (\tau) $
as follows:
\begin{equation}
l_0 ={\cal Z}_{ l} ( \tau) \, l (\tau) \,\, \,\,; \,\,\,\,
\theta_0 =\theta (\tau) + {\cal Z}_{\theta} (\tau) \, ,
\label{bare-renormalized}
\end{equation}
where $ {\cal Z}_{ l} $ and $ {\cal Z}_{\theta} $ are
renormalization constants for multiplicative amplitude
renormalization and additive phase renormalization respectively.
They are given by
\begin{equation}
{\cal Z}_l (\tau) = 1+ a (\tau) + \cdot\cdot\cdot   \,\,\,\,  ,
\,\,\,\,
 {\cal Z}_{\theta} (\tau) = b (\tau) +\cdot\cdot\cdot \, .
 \label{renormalizationconst}
 \end{equation}
 The $ \cdot\cdot\cdot $ means the terms to be involved while the
 approximation under consideration goes beyond the small displacement approximation.
 Substituting Eqs.(\ref{bare-renormalized}) and
(\ref{renormalizationconst}) into Eq.(\ref{baresol}) leads us to
choose
\begin{equation}
a(\tau)= \frac{A }{ 720 \pi^2 m}  \, \omega_0^4 \, ( \tau-t_0) \,
\,\,\, , \,\,\,\,
 b( \tau) = \frac{A }{ 240
\pi^2  m} \, \omega_0^2 \, \Lambda \, ( \tau-t_0) \, ,
\end{equation}
so as to  remove the secular terms containing $ \tau-t_0$. After
doing  renormalization, the solution is given by
Eq.(\ref{baresol}) as $ l_0, \theta_0$ and $t_0$ are replaced by $
l (\tau), \theta (\tau) $ and $ \tau$ respectively. The
independence  of the time scale $ \tau$ on $ l_0 $ and $ \theta_0
$ can lead to the renormalization group equations by taking the
$\tau$ derivative on Eq.( \ref{bare-renormalized}), which are of
the form:
\begin{eqnarray}
\frac{d}{d \tau} l (\tau) &=& - \frac{ A}{ 720 \pi^2
 m} \, \omega_0^4 \, l (\tau) \, , \nonumber \\
 \frac{d}{d \tau} \theta (\tau) &=& - \frac{A }{ 240 \pi^2  m} \, \omega_0^2 \,  \Lambda \, ,
\end{eqnarray}
with the solutions:
\begin{equation}
l (\tau)= l_0 \, e^{- \frac{A}{720 \pi^2 m} \, \omega_0^4 \, (
\tau-t_0)} \,\,\,\, , \,\,\,\, \theta (\tau) =\theta_0 -
\frac{A}{240 \pi^2 m} \, \omega_0^2 \, \Lambda \, ( \tau-t_0) \, .
\end{equation}
A change of the renormalization point $\tau$ is compensated by a
change in the renormalized amplitude  $ l (\tau)$ and phase  $
\theta (\tau)$. Substituting the solutions above to the
renormalized solution and setting $ \tau=t$, we obtain
\begin{equation}
\delta q(t) = l_0 \, e^{- \frac{A}{720 \pi^2 m} \, \omega_0^4 \, t
} \left\{ \cos \left[ \, \omega_0 \, ( 1+ \frac{A}{ 240 \pi^2 m}
\, \omega_0^2 \Lambda \, ) [ \, t \,-\, \theta_0 \, (
1-\frac{A}{240 \pi^2 m} \, \omega_0^2 \, \Lambda \, ) \, ] \right]
\right\} \, ,
\end{equation}
where the initial time has been set at $t_0=0$.

Obviously, the term of  forth time derivative in
Eq.(\ref{meanlangevin}) modifies the dispersive part of the mirror
by changing the oscillation frequency as well as shifting the
phase. The relaxation time scales are mainly determined from the
term of fifth time derivative given by
\begin{equation}
t_{\rm relax} \simeq  720 \pi^2 \, \left(  \frac{ c^2 }{A \,
\omega_0^2 } \right) \, \left( \frac{ m  c^2}{ \hbar  \omega_0 }
\right)
 \frac{1}{\omega_0} \, .
\end{equation}
   $ mc^2
>> \hbar \omega_o $ holds  for a macroscopic
mirror. Typically, the  emitted quanta driven by a nonuniform
accelerated mirror  is with a frequency which is  the same as the
oscillation frequency of a mirror. This condition means that the
energy loss from emitted quanta is far much less than the  rest
mass energy of a microscopic mirror. Thus, the recoiled effect of
the mirror for this process is small where one can provide  a
prescribed motion of the mirror, and  then find its correction
arising from the effects of quantum fields. The validity of the
small displacement approximation imposes the condition of $ l_0
\omega_0 << 1 $. Then, the order of magnitude of the relaxation
time scales can be obtained as
\begin{equation}
t_{\rm relax}
>> 10^4 \left( \frac{c} { l_0 \, \omega_0 } \right)^2
 \frac{1}{\omega_0} >> 10^4 \frac{1}{\omega_0} \, ,
\end{equation}
where $ A \approx l_0^2 $ has been assumed.  Thus, the very long
time scales for  having at least much more than  $ 10^4 $
oscillations are  needed to detect tiny damping on  the amplitude
of the oscillating mirror~\cite{FV}.

We now study the  fluctuations effects from quantum fields on the
mirror. The vacuum fluctuations are of great importance in early
times, say $ t<< t_{\rm relax}$, as the  dissipation effects  can
be ignored. The equation of the mirror then reduces to
\begin{equation}
m \,  \delta \ddot{q} (t) + m \omega_0^2 \, \delta q(t) = \eta (t)
\, .
\end{equation}
Its solution is obtained as
\begin{equation}
\delta v (t) = \int_0^t dt' \, \cos\left[ \omega_0 (t-t') \right]
\frac{ \eta(t')}{m} \, ,
\end{equation}
leading  to the  velocity fluctuations given by
\begin{eqnarray}
\Delta \delta v^ 2 (t) &=& \langle \delta v^2 (t) \rangle -\langle
\delta v(t)
\rangle^2  \nonumber \\
&=& \frac{1}{m^2} \int_0^t dt_1 \int_0^t dt_2 \cos \left[ \omega_0
(t-t_1) \right] \cos \left[ \omega_0 (t-t_2) \right] \left[
\langle \eta(t_1) \eta (t_2) \rangle -\langle \eta(t_1) \rangle
\langle\eta (t_2) \rangle \right] \nonumber \\
&=& \frac{1}{360 \pi^2} \frac{A}{m^2} \int_0^t dt_1 \int_0^t dt_2
\int_0^{\infty} \frac{ d \omega }{2 \pi} \, \omega^5 \,  \cos
\left[ \omega_0 (t-t_1) \right] \cos \left[ \omega_0 (t-t_2)
\right] \cos\left[ \omega (t_1-t_2) \right] \, , \nonumber \\
\label{vv-vac}
\end{eqnarray}
where we have used the fact that the forces from vacuum
fluctuations are Gaussian with correlations given by
Eq.(\ref{ff-vac-1}).
 We change variables of integration as $
u=t_1-t_2, v=t_1+t_2$, and the integral  above in terms of $ u,v$
is of the form :
\begin{eqnarray}
\Delta \delta v^2 (t)&=& \frac{A}{1440\pi^2 m^2} \,
\int_0^{\infty} \frac{ d\omega}{2 \pi}  \,\omega^5 \,
  \left\{
\int^0_{-t} du \int_{-u}^{u+2t} + \int^t_{0} du \int_{u}^{2t-u}
\right\} \,  \nonumber\\
&& ~~~~~~~~~~~~~~~~~~~~~~~~~~ \left\{ \,\cos \left[ \,\omega_0\,
(2t-v) \, \right] + \cos \left[ \, \omega_0 \, u\, \right]
\right\} \cos \left[\, \omega \, u\, \right] .
\end{eqnarray}
For the time $t$, say $ 1/\omega_0 << t << t_{\rm relax} $, we
find that the velocity fluctuations grow linearly in $t$ as
\begin{eqnarray}
\Delta \delta v^2 (t) && \simeq \frac{A}{720\pi^2 m^2} \, t\,
\int_0^{\infty} \frac{d \omega}{ 2\pi} \, \omega^5 \,
\int_0^{\infty} du \,  \left[ \cos \left[ (\omega+\omega_0 ) u
\right] +\cos
\left[ (\omega- \omega_0 ) u \right] \right] \,  \nonumber\\
&& \simeq \frac{A}{720\pi^2 m^2} \, t\,\int_0^{\infty} \frac{d
\omega}{ 2\pi}\, \omega^5 \, \pi \left[ \delta (\omega+\omega_0)
+ \delta (\omega-\omega_0) \right] \, \nonumber\\
&& \simeq \frac{A}{1440\pi^2 m^2} \,\omega_0^5 \,  t \, .
\end{eqnarray}
It can be seen that the typical frequency of quanta  absorbed by
the moving mirror to increase its velocity fluctuations  is the
frequency of the oscillating mirror. The energy gained from vacuum
fluctuations for each oscillation  can be obtained as
\begin{equation}
E  \simeq  \frac{1}{1440 \pi^2 } \left(  \frac{ A \, \omega_0^2
}{c^2} \right) \left( \frac{ \hbar \omega_0 }{ m c^2} \right) \,
\hbar \omega_0 \, ,
\end{equation}
with the order of magnitude given by
\begin{equation}
E  << \,  10^{-4} \, \left( \frac{l_0 \, \omega_0}{c}\right)^2 \,
\hbar \omega_0 << \, 10^{-4} \, \hbar \omega_0 \, ,
\end{equation}
where again $ A \approx l_0^2 $ and $ m c^2 >> \hbar \omega_0$
have been used. Thus, roughly about fewer than $ 10^{-4}$ quanta
with frequency $ \omega_0$  is absorbed by a mirror per
oscillation per area $ l_0^2$. Thus, the effects from vacuum
fluctuations can hardly be detected. The largely nonuniform
acceleration of a microscopic object can possibly amplify vacuum
fluctuations  where the treatment to tackle this issue beyond the
small displacement approximation is required.

\subsection{Thermal fluctuations}

This section will be devoted to understanding the dynamics of
moving mirrors in thermal fields. The large time and high
temperature limits give rise to $ \mid t-t' \mid
>> l
>> \tau_{\rm B} $  where $ A= \pi l^2$,  area of the mirror,
 and $ \tau_{\rm B} \equiv 1/( \pi k_{\rm B} T)
$, a characteristic thermal correlation length scale. Then, the
Green's function for scalar fields in Eqs.(\ref{freegreen}) and
(\ref{reimgreen}) can be approximated by:
\begin{equation}
G^{ ( >,< )} ( {\bf x}-{\bf x'} ; t-t') \simeq \frac{1}{ 4\pi^2
\tau_{\rm B} \mid {\bf x}-{\bf x'} \mid}  \left[ e^{ -\frac{2}{
\tau_{\rm B}} (\mid t-t'\mid + \mid {\bf x}- {\bf x'} \mid )}-
e^{- \frac{2}{ \tau_{\rm B}} (\mid t-t'\mid - \mid {\bf x}- {\bf
x'} \mid )} \right] \, .
\end{equation}
Thus, the force correlations including thermal effects can be
obtained from Eqs.(\ref{anticommutator}) and (\ref{ffcorrelation})
as:
\begin{eqnarray}
\sigma_{FF} (t-t') & \simeq & \frac{ 16 \, l^2}{ \pi^2 \tau_{\rm
B}^6 } \, \left\{ \left( 1 +  \frac{1  }{  4 \left( \frac{l}{
\tau_{B}} \right) } - \frac{1}{ 32 \left( \frac{l}{\tau_{\rm B}}
  \right)^4 }      \right)
  e^{ -\frac{4}{\tau_{\rm B}} \mid t-t'\mid } \,  - \left( \frac{1}{ 16 \left(
  \frac{l}{ \tau_{\rm B}} \right)^3 } -\frac{1}{ 64 \left( \frac{l}{\tau_{\rm B}}
  \right)^4 } \right) \, e^{ -\frac{4}{\tau_{\rm B}} ( \mid t-t'\mid
  -l) } \right. \nonumber \\
  &+ & \left. \left( \frac{1}{ 16 \left(
  \frac{l}{ \tau_{\rm B}} \right)^3 } +\frac{1}{ 64 \left( \frac{l}{\tau_{\rm B}}
  \right)^4 } \right) \, e^{ -\frac{4}{\tau_{\rm B}} ( \mid t-t'\mid
  +l) }
\right\}   \, \nonumber  \\
& \simeq & \frac{ 16 \, l^2 }{ \pi^2  \tau_{\rm B}^6 } e^{
-\frac{4}{\tau_{\rm B}} \mid t-t'\mid } \, ,
\end{eqnarray}
 which can be further  approximated by
\begin{equation}
\sigma_{FF} (t-t') \simeq \frac{ 8 \, l^2}{ \pi^2 \tau_{\rm B}^5}
\, \delta (t-t')
\end{equation}
using the fact that
\begin{equation}
\lim_{\alpha \rightarrow \infty} \frac{\alpha }{2} \,  e^{ -
\alpha \mid x \mid } = \delta (x) \, .  \label{deltafun}
\end{equation}
It reveals  that the high temperature fluctuations are of
uncorrelated white noise. Using Eq.(\ref{df-high-T}), the
fluctuation-dissipation theorem in the high-T limit, one can
determine the imaginary part of $\chi_{FF} (\omega)$ that leads to
the dominant effect on dissipation with the term proportional to
the mirror's velocity. It is due to the force fluctuations. The
real part of $\chi_{FF} (\omega) $ renormalizes the oscillation
frequency as well as the mass of the mirror with temperature
corrections. However, the corresponding temperature correction to
the oscillation frequency, which describes a position dependent
static force, vanishes since the mean pressure force from thermal
scalars on the mirror is zero by the symmetry argument~\cite{JR2}.
The mass will acquire the temperature correction which is
subdominant as its correction is suppressed by a factor of $ \hbar
\omega_0 / k_{\rm B} T $  comparing with the damping term. From
Eq.(\ref{motioninduced-kernel}), the high-$T$  motion-induced
force can be obtained from the corresponding dissipative force,
and is found to be also proportional to the mirror's velocity. It
arises from the Doppler shift of thermal scalars.

 Thus, involving the dominant thermal effects,
 the Langevin equation now becomes:
\begin{equation}
m \, \delta \ddot{q} (t) +  \gamma_{T} \, \delta \dot{q} (t) + m
\omega^2_0 \, \delta q (t) = \eta (t) \,
\end{equation}
 with the white noise correlations:
\begin{equation}
\langle \eta(t) \eta(t') \rangle = 8  \pi^2 c^3 A  \, \left(
\frac{k_{\rm B} T}{ \hbar c} \right)^3  \, \left( \frac{ k_{\rm B}
T}{ c^2} \right)^2 \, \delta (t-t') \, . \label{ff-T-1}
\end{equation}
The damping coefficient can be found to be~\cite{MMF}:
\begin{equation}
\gamma_{T} \simeq  8 \, \pi^2   c A  \, \left( \frac{k_{\rm B} T}{
\hbar c} \right)^3  \, \left( \frac{ k_{\rm B} T}{ c^2} \right) \,
.
\end{equation}
 The relaxation time scales, $ t_{\rm relax} \simeq (\gamma_{T} /
m)^{-1} $, are  the time scales when dissipation effects become
important. To obtain the maximal fluctuations for the mirror, we
now consider the time scales, say $ t_{\rm relax}
>> t
>> l $, where dissipation effects can be ignored. We find that
\begin{equation}
\Delta \delta v^2 (t)  \simeq 4 \pi^2 c^3 A \, \left( \frac{k_{\rm
B} T}{ \hbar c} \right)^3  \, \left( \frac{ k_{\rm B} T}{ m c^2}
\right)^2 \,
 \, t \, .
\end{equation}
The maximal velocity fluctuations  can be achieved  by roughly
setting the time scales, $ t= t_{\rm relax} $,  as follows:
\begin{equation}
\Delta \delta v^2_{\rm max} (t) \simeq c^2 \, \left( \frac{ k_{\rm
B} T}{ m c^2} \right) \, .
\end{equation}
Thus, it leads to
\begin{eqnarray}
\frac{\Delta l_{\rm max}}{ l_0 } & \simeq & \frac{ \Delta \delta
v_{\rm max} }{ \delta v } \simeq \left( \frac{c}{ l_0 \, \omega_0}
\right) \, \left( \frac{ k_{\rm B} T}{ m c^2}
\right)^{\frac{1}{2}} \,
\nonumber \\
& \simeq & 10^{-8} \, \left( \frac{ 10 \, {\rm cm}}{ l_0} \right)
\, \left( \frac{1 \, {\rm s}^{-1} }{ \omega_0} \right) \, \left(
\frac{1 \, {\rm kg}}{ m} \right)^{\frac{1}{2}} \, \left( \frac{T}{
1 \, {\rm kev }} \right)^{\frac{1}{2}}
 \,
\end{eqnarray}
with the corresponding relaxation time scales given by
\begin{equation}
t_{\rm relax} \simeq 10^{-2} {\rm s} \left( \frac{ 100 \, {\rm
cm}^2}{A} \right) \, \left( \frac{m}{ 1 \, {\rm kg}} \right) \,
\left( \frac{ 1\, {\rm kev}}{ T} \right)^4 \, ,
\end{equation}
where $l_0$ and $ \omega_0$ are the typical  oscillation amplitude
and frequency of the mirror.   As long as the temperature of
thermal fields is of order $ {\rm kev}$, the amplitude
fluctuations of the oscillating mirror are of order $ 10^{-8} \,
l_0 $ within the time scales of $ 10^{-2} \, {\rm s}$, which can
be detectable. The mass correction from thermal effects can be
obtained from Eq.(\ref{massrenormalized}) by replacing the energy
cutoff $ \Lambda $ with the typical thermal energy $ k_{\rm B} T$
given by~\cite{MMF} :
\begin{equation}
\Delta m_{T} \simeq - A \left( \frac{ k_{\rm B} T}{ \hbar c}
\right)^2 \, \left( \frac{k_{\rm B} T }{ m c^2} \right ) \, m
\simeq -10^{-16} m
\end{equation}
with the above value of the parameters.  This extremely small mass
correction can be ignored in our calculations.

\section{ Conclusions}

In this paper, we present a general framework for describing the
dynamics of  moving mirrors  in quantum fields in the case where the mirror undergoes the
small displacement.    The mirror of perfect reflection imposes
the boundary conditions on field fluctuations, and  leads to the
coupling between the mirror and  fields. The force on the mirror
is given by the area integral of the stress tensor of the fields.
Using the Schwinger-Keldysh formalism, coarse-graining quantum
fields leads to the stochastic behavior in the mirror's trajectory
encoded in the  coarse-grained effective action with the method of
influence functional. In the semiclassical regime, the Langevin
equation can be derived  involving backreaction effects. We find
that the Langevin equation reveals two levels of backreaction
effects on the dynamics of the mirror:  radiation reaction induced
by the motion of the mirror as well as backreaction dissipation
arising from fluctuations of quantum fields via a
fluctuation-dissipation relation. The corresponding
fluctuation-dissipation theorem is derived for quantum fields in
vacuum and at finite temperature respectively. We find that,
although the theorem of fluctuation and dissipation for the case with the
 small mirror's displacement is of model independence, the obtained theorem
from the  first principles derivation reveals that it is also {\it
independent } of the regulators introduced to deal with
short-distance divergences from quantum fields. Thus, when the
method of regularization is introduced to compute the dissipation
and fluctuation effects, this theorem must be fulfilled as the
results are obtained by taking the short-distance limit in the end
of calculations. This theorem also allows us to compute the
dissipation kernel from the obtained fluctuation kernel  and vice
versa.

Consider  a  situation where the mirror is attached to a spring
and undergoes oscillations with a natural frequency $ \omega_0$.
In vacuum, we find that the relaxation time scales for having much
more than $ 10^4 $  oscillations are needed to detect tiny damping
on  the oscillation amplitudes of the mirror due to the
backreaction effects. The energy gain of the mirror from vacuum
fluctuations is by absorbing fewer than $ 10^{-4}$ quanta for each
oscillation with frequency $ \omega_0 $. Thus, these vacuum
fluctuations can hardly be detected. The largely nonuniform
acceleration of a microscopic object can possibly amplify the
effects of vacuum fluctuations where the treatment to tackle this
issue beyond the small displacement approximation is required. On the contrary, at
finite temperature, as long as the temperature of thermal fields
is of order $ {\rm kev}$, the ratio of the amplitude fluctuations
to the amplitude  of the oscillating mirror are of order $ 10^{-8}
$ within the time scales of $ 10^{-2} \, {\rm s}$, leading to the
detectable effects.

\begin{acknowledgments}
We would like to thank Bei-Lok Hu and Larry H. Ford for
stimulating discussions.  Part of this work was done when DSL
visited the Theory Group at the Blackett Laboratory, Imperial
College in London, U. K.. DSL would like to thank  R. J. Rivers
and T. S. Evans for their hospitality. This work was supported in
part by the National Science Council, R. O. C. , under grant
NSC92-2112-M-259-007.
\end{acknowledgments}


\end{document}